\documentclass[preprint,review,12pt,sort&compress]{./template/elsarticle/elsarticle}
\usepackage{epstopdf}
\usepackage{amsmath}
\usepackage{amsfonts}
\usepackage{xcolor}
\usepackage{lineno,hyperref}
\usepackage{placeins} % FloatBarrier
\usepackage{array} % column-centered text in tables
\usepackage{mathtools}
\usepackage{rotating} % sidewaysfigure
\modulolinenumbers[5]
\pdfoutput=1

\journal{Combustion and Flame}
\newcommand*\diff{\mathop{}\!\mathrm{d}} % integral d
\newcommand{\sm}[1]{{\scriptscriptstyle#1}} % make number in sub/superscript same size as text
\newcommand{\eqdef}{=\vcentcolon} % =: (mathtools)
\newcolumntype{P}[1]{>{\centering\arraybackslash}p{#1}}
\renewcommand\labelitemi{\quad--}
\newcommand{\tabitem}{~~\llap{\labelitemi}~~}
\begin{document}

\begin{frontmatter}

\title{The Role of Differential Diffusion during Early Flame Kernel Development under Engine Conditions - Part~I: Analysis of the Heat-Release-Rate Response}

%% Group authors per affiliation:

%% or include affiliations in footnotes:
\author[mymainaddress]{Tobias Falkenstein}
%\ead{office@itv.rwth-aachen.de}

\author[mymainaddress]{Aleksandra Rezchikova}
%\ead{office@itv.rwth-aachen.de}

\author[mymainaddress]{Raymond Langer}
%\ead{office@itv.rwth-aachen.de}

\author[mymainaddress]{Mathis Bode}
%\ead{office@itv.rwth-aachen.de}

\author[mysecondaryaddress]{Seongwon Kang}
%\ead{skang@sogang.ac.kr}

\author[mymainaddress]{Heinz Pitsch\corref{mycorrespondingauthor}}
\cortext[mycorrespondingauthor]{Corresponding author}
\ead{office@itv.rwth-aachen.de}

\address[mymainaddress]{Institute for Combustion Technology, RWTH Aachen University, 52056 Aachen, Germany}
\address[mysecondaryaddress]{Department of Mechanical Engineering, Sogang University, Seoul 121-742, Republic of Korea}

\begin{abstract}
% 100 - 300 words
Although experimental evidence for the correlation between early flame kernel development and cycle-to-cycle variations (CCV) in spark ignition (SI) engines was provided long ago, there is still a lack of fundamental understanding of early flame/turbulence interactions, and accurate models for full engine simulations do not exist. Since the flame kernel is initiated with small size, i.e. with large positive curvature, differential diffusion is expected to severely alter early flame growth in non-unity-Lewis-number (${\mathrm{Le}\neq1}$) mixtures as typically used in engines. In this work, a DNS database of developing iso-octane/air flame kernels and planar flames has been established with flame conditions representative for stoichiometric engine part-load operation. Differential diffusion effects on the global heat release rate are analyzed by relating the present findings to equivalent flames computed in the ${\mathrm{Le}=1}$ limit. It is shown that in the early kernel development phase, the normal propagation velocity is significantly reduced with detrimental consequences on the global burning rate of the flame kernel. Besides this impact on the overall mass burning rate, the initial production of flame surface area by the normal propagation term in the flame area balance equation is noticeably reduced. 
By using the optimal estimator concept, it is shown that strong fluctuations in local heat release rate inherent to ${\mathrm{Le}\neq1}$ flames  in the thin reaction zones regime are mainly contained in the parameters local equivalence ratio, enthalpy, and H-radical mass fraction. Differential diffusion couples the evolution of these parameters to the unsteady flame geometry and structure, which is analyzed in Part~II of the present study~(Falkenstein et al., Combust. Flame, 2019\nocite{Falkenstein19_kernel_Le_II_cnf}).
%To address future modeling efforts, other parameter choices  have been quantitatively tested and related to the ${\mathrm{Le}=1}$ limit, which features very small variations in heat release rate. 
%Finally, the effect of the particular flame-kernel-curvature-distribution-shape caused by characteristic small kernel/turbulence interactions is shown to have a minor effect on the mean heat release rate in comparison to the reduction by mean positive kernel curvature. \par
%
\end{abstract}

\begin{keyword}
Flame Kernel \sep Differential Diffusion \sep Flame Stretch \sep DNS \sep Premixed Flame \sep Spark-Ignition Engine
\end{keyword}

\end{frontmatter}

%\linenumbers
%============================================================================================================================================
\section{Introduction}
\label{sec:intro}
Combustion stability  is a prerequisite for more efficient spark ignition (SI) engine operation~\cite{Aleiferis04_lean_ccv,Jung17_lean_si_ccv} and reduced engine-out emissions~\cite{Milkins74_CO_HC_ccv_exp,Karvountzis17_emissions_ccv}. 
%Even in homogeneous-charge, port-fuel-injection engines, 
The occurrence of cycle-to-cycle variations~(CCV) is mainly attributed to the very early combustion phase~\cite{Drake07_si_engine_devel}, which may take up to 30\,\% of the entire combustion duration to oxidize less than 2\,\% of the in-cylinder fuel mass~\cite{Maly94_early_comb_phase_dur}. By using advanced laser-optical diagnostics, the time until the young laminar flame kernel transitions to a turbulent flame (at approx. 1\,\% burned mass) was shown to correlate with CCV~\cite{Schiffmann17_exp}. Further, it was found that stretch effects in ${\mathrm{Le}\neq1}$ mixtures affect the duration until 10\,\% of fuel are consumed, which correlates with CCV as well. Although differential diffusion effects have been systematically investigated in spherical expanding turbulent flames by several experimental research groups~\cite{AbdelGayed85_Exp_Le_Turb,Aung02_exp_turb_flame_pref_diff,Saha14_Markstein_effect_flame_surf,Brequigny16_Ma_effect_turb_exp}, limited accessibility of the smallest time- and length scales demands for complementary numerical analyses. \par
% Drake & Haworth:  https://www.sciencedirect.com/science/article/pii/S154074890600383X?via%3Dihub
% Homogeneous-charge spark-ignition port-fuel-injection (PFI) engines
% Ignition and early flame-kernel growth dominate cycle-to-cycle variability in burn rate.
%
% Although the transition from spark to fully developed turbulent flame propagation corresponds to only a few percent
% of the total fuel burned, it typically represents about 30% of the total burn duration in an engine cycle [13] and 
% accounts for a large fraction of the cycle-to-cycle variability in combustion performance.
%
While \textit{flame kernels} developing in turbulent flow fields have been computed in various DNS studies~\cite{Jenkins02_dns_3d,Thevenin05_kernel_dns_3d,Klein06_kernel_dns_stretch,Jenkins06_kernel_dns_stretch,Chakraborty07_spark_ign_turb_dns,Chakraborty07_kernel_stretch,Bastiaans09_kernel_dns_3d,Chakraborty09_kernel_dns_3d_fsd,Shalaby10_kernel_dns_3d}, differential diffusion effects have been considered mainly in recent years~\cite{Echekki94_kernel_dns_Le_effect,Dunstan09_kernel_H2_enrich,Fru11_kernel_dns_3d,Pera13_kernel_dns_3d,Yenerdag15_H2_kernel_dns,Dinesh16_spherical_H2_dns,Uranakara17_ign_kernel_3d,Saito18_CH4_C7H16_kernel_ign,Alqallaf19_kernel_curv_eq}. A non-exhaustive list of DNS parameters and flame conditions that have been investigated to date is provided in Tab.~\mbox{S-1} of the supplementary material. \par %\ref{tab:kernel_dns_summary}. \par
In the early DNS study by Echekki et al.~\cite{Echekki94_kernel_dns_Le_effect}, a Lewis number variation was performed for the very first time in a flame kernel configuration. It was found that flame growth is significantly accelerated in a mixture with ${\mathrm{Le}=0.53}$ compared to ${\mathrm{Le}=1}$ conditions. More detailed insights were enabled only many years later by increasing computational power and availability of high-performance computing resources. Dunstan and Jenkins~\cite{Dunstan09_kernel_H2_enrich} investigated the effect of hydrogen enrichment on the development of lean methane/air flame kernels, in analogy to a previous planar-flame study by Hawkes and Chen~\cite{Hawkes04_focusing}. The behavior of pure methane/air flames was shown to be similar in both configurations. By contrast, hydrogen enrichment yielded a stronger enhancement of the flame kernel turbulent burning velocity due to the stronger impact of reduced thermal-diffusive stability in presence of higher global stretch rate. Dinesh et al.~\cite{Dinesh16_spherical_H2_dns} performed DNS of lean $\mathrm{H}_{2}$/$\mathrm{CO}$/air (syngas) flame kernels at two Reynolds numbers with  ${\mathrm{Le}=0.385}$ (realistic) and ${\mathrm{Le}=1}$. At elevated turbulence levels, the significant increase in flame area caused by thermal-diffusive effects observed at the lower Reynolds number was found to diminish. Differential diffusion was shown to increase the local burning velocity, hence accelerating flame kernel growth, even at the higher Reynolds (and moderate Karlovitz) number. Alqallaf et al.~\cite{Alqallaf19_kernel_curv_eq} systematically varied the Lewis number ($\mathrm{Le}=0.8,1.0,1.2$) in DNS of flame kernel development computed with single-step chemistry to analyze the effect on flame curvature dynamics. Flame propagation was shown to smoothen the flame surface for ${\mathrm{Le}\geq 1}$, while production of both positive and negative curvatures is promoted in mixtures with ${\mathrm{Le}< 1}$. \par
Extinction of an  $\mathrm{H}_2$/air ignition kernel, i.e.\ before the critical radius for self-sustaining flame propagation is reached, was studied by Uranakara et al.~\cite{Uranakara17_ign_kernel_3d}. A particle-based analysis was utilized to understand the effect of turbulence-enhanced heat loss leading to extinction. Saito et al.~\cite{Saito18_CH4_C7H16_kernel_ign} performed two-dimensional DNS to study the effect of small-scale turbulence on the ignition of methane/air and n-heptane/air flame kernels. After normalization with the laminar flame and ignition delay times, similar sensitivities to strain were observed for both mixtures. \par
Regarding DNS studies on \textit{developed} turbulent flames, those which were performed under engine-relevant conditions or those which considered larger hydrocarbon fuel species are particularly relevant for the present work. Savard et al.~\cite{Savard17_si_engine_dns} analyzed effects of pressure on turbulent, slightly lean (${\phi = 0.9}$) iso-octane/air flames at different Karlovitz numbers. Laminar chemical reaction pathways primarily contributing to heat release were significantly changed by ambient pressure. It was found that at low pressures of 1\;bar, strong variations in heat release were caused by higher sensitivity to curvature due to differential diffusion as compared to 20\;bar. Increased Karlovitz numbers lead to reduced fluctuations in heat release due to enhanced turbulent species transport inside the flame structure. \par
% Note that Wang2019 observed strong local equiv. ratio variations in H2 flames at 20bar, he attributes it to thinner reaction zones at high pressure (which is different from Savard's kinetic arguments at const. Ka.)
% our 6bar pressure case is relevant for engine part-load. Ka low in conventional engine at stoichiometric conditions. (Savard slightly lean and higher Ka, lt=lf) -> stretch effects can be significant here
Aspden et al.~\cite{Aspden17_Le_effect_C12H26} computed lean dodecane/air flames for a range of Karlovitz numbers (${\mathrm{Ka}=}$1-36) in the thin reaction zones regime. Small-scale turbulent mixing was found to cause significant deviations in reaction rates from laminar flames computed with ${\mathrm{Le}=1}$ and ${\mathrm{Le}>1}$, which is in contrast to methane/air flames~\cite{Aspden16_ch4_flame}. In the low-temperature region of the dodecane/air flame structure, reaction rates of fuel and intermediate species were shown to be significantly reduced at elevated Karlovitz numbers, which was attributed to turbulent perturbations of the fuel breakdown. This effect was identified as a distinct feature of heavy-hydrocarbon flames due to the spatial separation of fuel consumption and heat release. \par %The negative correlation between burning velocity and curvature due to differential diffusion did suppress flame wrinkling. \par
%
% prod. rates CH4, Ka=36: For all of these species, the net production is essentially unchanged from the laminar flame solution.
% prod. rates C12H26, Ka=36: Furthermore, unlike methane, the peak rate for each species is dramatically reduced in
%  the turbulent case  relative to the laminar flame. The response to turbulence is distinctly different
%  than the corresponding unity Lewis number laminar flame as indicated by  the blue curve in Fig. 10.
%  This suggests  that the turbulence is beginning to disrupt the chemistry in the cool portion of the
%  flame where the dodecane is being consumed.
%
% Compared to methane, it was found that because the fuel decomposition pyrolysis and oxidation occurs in cooler
% regions of the flame, the resulting distribution of fuel fragments tends to be susceptible (anfaellig) to mixing
% processes. The basic oxidation pathways of these species is largely unaffected by this process, but the magnitudes
% are considerably reduced. Consequently, and despite the disruption due to turbulence..
%
Although significant progress has been made towards more realistic DNS of premixed flames using detailed chemistry, investigations on the behavior of actual road transportation fuels under engine-relevant turbulence and thermodynamic conditions are still very scarce. While few studies on canonical planar flame configurations exist in literature, the role of molecular transport, as well as its interaction with small-scale turbulent mixing and chemistry inside the flame structure of flame kernels has not been studied in a realistic setting. Experimental evidence on the relevance of early flame kernel development for~CCV in~SI engines and the lack of accurate models suggest to address this gap in literature. {\color{black}To this end, a new~DNS database was designed to be representative for SI~engine part-load conditions. Systematic parameter variations were conducted that facilitate the isolation of effects related to the small flame kernel size respective of the hydrodynamic length scales, as well as of differential diffusion effects. The database consists of five different flame configurations, which have been partly analyzed in previous studies as shown in Tab.~\ref{tab:dns_datasets}. To systematically approach the complexity of early flame kernel development, the overall analysis has been divided into four parts by sequentially considering unity- and non-unity Lewis number flames, as well as macroscopic and micro-scale effects. Run-to-run variations in the global heat release rate of flame kernels  computed in the ${\mathrm{Le}=1}$ limit were attributed to the effect of flame front/turbulence interactions on flame surface area evolution~\cite{Falkenstein19_kernel_Le1_cnf}. In a subsequent study, the investigation of ${\mathrm{Le}=1}$ flames with different $D_{0}/l_{\mathrm{t}}$ (ratio of the initial flame diameter and the integral length scale of turbulence) has shown that such variations in the total flame area are caused by stochastic variations in curvature variance due to large-scale turbulent flow motion with characteristic length scales of at least the flame kernel size~\cite{Falkenstein19_kernel_Le1_jfm}. The present study consists of two parts, where the ${\mathrm{Le}>1}$ datasets are analyzed in detail. 
\begin{table}%[!htb] %%%%%%%%%%%%%%%%%%%%%%%%%%%%%
\centering
{\color{black}
  \caption{Flame realizations available in the overall DNS database.}
\vspace{0.1cm}
%\begin{tabular}{P{.22\textwidth}|P{.22\textwidth}}
\begin{tabular}{P{.3\textwidth}|P{.12\textwidth}|P{.12\textwidth}|P{.12\textwidth}} %% review_12pt
\hline
%~ & ~ & ~ & ~ \\ [-10pt]
%~ & ~ & \multicolumn{2}{c}{Realizations} \\
~ & ~& \multicolumn{2}{c}{Number of Cases}  \\
Configuration &  $D_{0}/l_{\mathrm{t}}$  & $\mathrm{Le}=1$  & $\mathrm{Le}>1$ \\
~ & ~ & ~ & ~ \\ [-10pt]
\hline
~ & ~ & ~ & ~ \\ [-10pt]
Engine flame kernel & 0.3 & 2 & \textbf{4} \\ %[3pt]
~ & ~ & ~ & ~ \\ [-10pt]
\hline
~ & ~ & ~ & ~ \\ [-10pt]
Large flame kernel & 2.0 & 1 & ~ \\ %[3pt]
~ & ~ & ~ & ~ \\ [-10pt]
\hline
~ & ~ & ~ & ~ \\ [-10pt]
Planar flame & $\infty$ & 1 & \textbf{1} \\ %[3pt]
~ & ~ & ~ & ~ \\ [-10pt]
\hline
\hline
~ & ~ & ~ & ~ \\ [-10pt]
Reference & {} & \cite{Falkenstein19_kernel_Le1_cnf,Falkenstein19_kernel_Le1_jfm} & \cite{Falkenstein19_kernel_Le_II_cnf}, \textbf{this} \\[10pt]
%~ & ~ & ~ & ~ \\ [-10pt]
\hline
\end{tabular}
\label{tab:dns_datasets}
}
\end{table} %%%%%%%%%%%%%%%%%%%%%%%%%%%%%
}
In the first part, the heat-release-rate response to differential diffusion effects is analyzed based on the integrated chemical source term, which has high practical relevance, and based on the local chemical source term, which is used to identify the governing parameters in ${\mathrm{Le}\neq1}$ flames in a quantitative fashion. In the second part~\cite{Falkenstein19_kernel_Le_II_cnf}, the coupling between the local mixture state, which determines the local heat release rate, and the flame geometry and structure is established. \par
%\textcolor{red}{The present work is a continuation of our previous study on flame kernel development under engine-relevant conditions in the ${\mathrm{Le}=1}$ limit~\cite{Falkenstein19_kernel_Le1_cnf,Falkenstein19_kernel_Le1_jfm}. The existing DNS database has been extended by several flame kernel realizations and a planar flame computed with realistic Lewis numbers. Hence, analyses with particular focus on the role of differential diffusion are enabled.} \textcolor{red}{[ToDo: comment on structure Part~I and~II] Investigations on early flame kernel development are supplemented by fundamental insights into the flame structure of a fully developed planar turbulent flame and are compared to laminar flamelets. In this way, the dominant effects that detrimentally reduce the global burning rate in ${\mathrm{Le}>1}$ flames can be identified. Finally, the impact of differential diffusion on the very early flame kernel growth is studied by considering the flame area evolution and revisiting the characteristic nature of flame kernel/turbulence interactions revealed in our previous work~\cite{Falkenstein19_kernel_Le1_jfm}.} \par
The present manuscript is organized as follows. In Sect.~\ref{sec:analyt_approach}, the overall analytical approach pursued to explore differential diffusion effects in the present engine-relevant datasets is summarized. A brief overview of the DNS database is provided in Sect.~\ref{sec:dns_database}. The global and local heat release rates are analyzed in Sect.~\ref{sec:results} and suitable parameters for a reduced representation of differential diffusion phenomena inside the flame structure are proposed.
\section{Analytical Approach}
\label{sec:analyt_approach}
% https://www.thwink.org/sustain/articles/000_AnalyticalApproach/index.htm
To assess the overall combustion process in technical combustion devices (e.g.\ in SI~engines), global parameters are typically of interest, such as the volume-integrated heat release rate or reaction progress variable source term:
\begin{equation}%%%%%%%%%%%%%%%%%%%%%%%%%%%%% 
\begin{aligned}
%\nonumber
\left.\overline{\dot{\;\omega_{c}}}\right|_{\Omega} & =  \frac{1}{V_{\Omega}} \int_{\Omega} \dot{\;\omega_c} \diff V .
\label{eq:integr_vol_srcProg}
\end{aligned}
\end{equation} %%%%%%%%%%%%%%%%%%%%%%%%%%%%%
Here, $c$ is a synonym for a reaction progress variable, e.g.\ a (normalized) temperature or a quantity representative for the major product species. In spark ignition engines, a fast burning rate is desirable mainly for three reasons. First, thermal efficiency is improved when approaching the limit of constant-volume combustion. Second, reduced residence times of the end gas (e.g.\ near hot surfaces) achieved by faster arrival of the flame front may reduce knock tendency. Third, an acceleration of the early flame kernel growth shortens the transition time to a fully developed turbulent flame~\cite{Maly94_early_comb_phase_dur,Schiffmann17_exp}. This makes the young flame less prone to stochastic kernel/flow interactions, which may significantly reduce CCV. However, differential diffusion effects may severely slow down the early flame kernel development for mixtures of common transportation fuels and air with ${\mathrm{Le}>1}$. \par
\begin{figure}
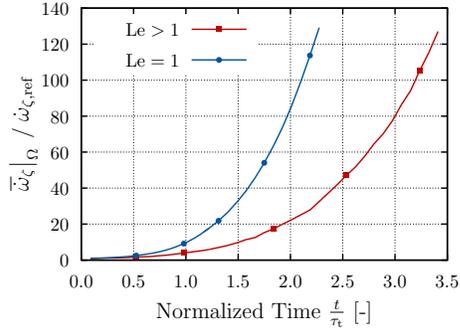
 %%%%%%%%%%%%%%%%%%%%%%%%%%%%%
\centering
\begin{minipage}[b]{0.45\textwidth}
  \graphicspath{{./data/FLAME_KERNEL_DNS_01/181003_curv_strain_kin_restor_diss_divg_skew/}}
  \centering
%trim={<left> <lower> <right> <upper>}
%\hfill\vspace{0.1cm}
  %\makebox[0pt][l]{\quad(a)}\includegraphics[trim={10cm -4cm 40cm -6cm},clip,width=5.5cm]{}
\input{template/change_font_10.tex} % review_12pt
  \scalebox{0.7}{\input{./data/FLAME_KERNEL_DNS_01/181003_curv_strain_kin_restor_diss_divg_skew/kernel_dns_mean_srcPROG_time_01_02_JFM_HALF.tex}}
\input{template/change_font_12.tex} % review_12pt
\end{minipage}
\caption{Effect of differential diffusion on the integrated progress variable source term for two equivalent flame kernel DNS datasets.}
\label{fig:integr_srcProg}
\end{figure}

% Due to this practical relevance, the effect of ${\mathrm{Le}>1}$  on the global burning rate of early flame kernels is assessed in this section. \par 
% check refs: https://www.sciencedirect.com/science/article/pii/S0360319913022945
% https://www.sciencedirect.com/science/article/pii/S0306261918312017
%To this end, the global burning rate is expressed as the domain-integrated reaction progress variable source term~$\left.\overline{\dot{\omega}_{\zeta}}\right|_{\Omega}$~(cf.\ Eq.~(\ref{eq:integr_vol_srcProg})), which is correlated with the heat release rate. 
To give a first impression of the impact of differential diffusion under engine conditions, the integrated reaction source term has been evaluated in a stoichiometric iso-octane/air flame kernel DNS dataset, and an equivalent dataset computed in the ${\mathrm{Le}=1}$ limit. In Fig.~\ref{fig:integr_srcProg} it is shown that ${\mathrm{Le}=1}$ overall leads to a substantially higher burning rate than in case of the realistic engine fuel, despite a 16\,\% lower \textit{unstretched} laminar burning velocity {\color{black}(cf.\ Tab.~\ref{tab:dns_mixture})}. To estimate the influence on~CCV in actual engines, the laminar-to-turbulent transition time~$\tau_{\mathrm{lam-turb}}$ proposed by Schiffmann et al.~\cite{Schiffmann17_exp} has been evaluated for both flames. It turns out that the presence of differential diffusion leads to a 31\,\% increase in~$\tau_{\mathrm{lam-turb}}$, which may increase CCV in terms of the coefficient of variance of the mean effective pressure~($\mathrm{COV}_{\mathrm{IMEP}}$) from 1.8 to 4.9\,\% (cf.\ Fig.~7 in~\cite{Schiffmann17_exp}). \par
% FLAME_KERNEL_DNS_01/180523_vol_area_size_I0_first/schiffmann_tau_lam-turb_02.ods
% FLAME_KERNEL_DNS_01/180523_vol_area_size_I0_first/schiffmann_tau_lam-turb_02_slide.pdf
% Spark timing in Paper: 342deg. Fig7: 348deg (w.r.t.ign=6deg)->COV=1.75%, 349.86=(342+6*1.31)->COV=4.87%
%Note that the results in Fig.~\ref{fig:integr_srcProg} are plotted as function of time, which is relevant for the engine application. However, this representation implies that differences in the integrated source term accumulate due to growing differences in flame kernel size. In order to rigorously analyze the role of differential diffusion during flame development, it is desirable to enable comparisons of data extracted from flames with similar geometry and size. Hence, results will be shown as function of kernel radius hereafter. The radius is here defined as the median of the distance distribution, measured between all flame points and the geometric center of the flame. \par
%
\begin{sidewaysfigure*}
%\begin{figure*} %%%%%%%%%%%%%%%%%%%%%%%%%%%%%
\centering
\begin{minipage}[b]{0.9\linewidth}
  \graphicspath{{./figures/flame_physics_schematic/inkscape/}}
  \centering
%trim={<left> <lower> <right> <upper>}
  %\includegraphics[trim={0cm 0cm 0cm 0cm},clip,width=\linewidth]{./figures/flame_physics_schematic/190418_pref_diff_flame_scheme_01-eps-converted-to.pdf}
\includegraphics[trim={0cm 0cm 0cm 0cm},clip,width=\linewidth]{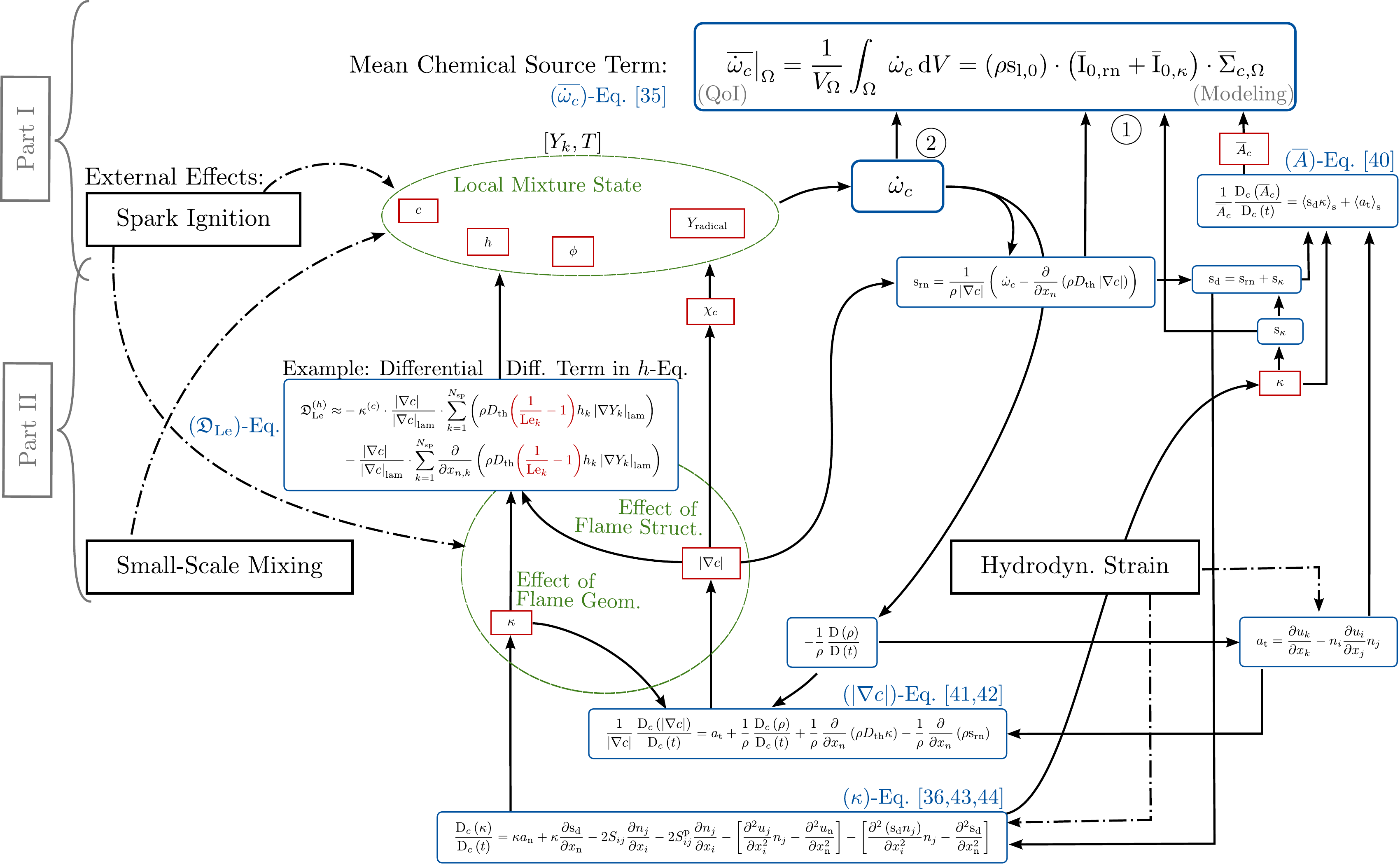}
% citations inside figure
\nocite{Falkenstein19_kernel_Le1_cnf} % global hrr eq.
\nocite{Candel90_fsd_eq,Falkenstein19_kernel_Le1_cnf} % total area eq.
\nocite{Kim07_sc_align,Sankaran07_sc_grad_eq} %,Dopazo15_sc_grad_eq,Wang17_jet_dns} grad-eq.
\nocite{Pope88_surf_turb,Dopazo18_curvEq,Falkenstein19_kernel_Le1_jfm} %,Cifuentes18_curv_eq_kempf,Alqallaf19_kernel_curv_eq} curv-eq.
%\nocite{Masi13_entropy_eq_chi,Borghesi15_entropy_eq_chi} % entropy eq.
% end citations
\end{minipage}
\caption{Reduced representation of turbulence/flame interactions on the integrated progress variable source term~$\overline{\dot{\;\omega_{c}}}$ in presence of differential diffusion (${\mathrm{Le}_k\neq1}$). Two analysis pathways are suggested: \textcircled{\tiny 1} A macroscopic perspective based on the FSD concept (cf. r.h.s.\ of \mbox{($\overline{\dot{\;\omega_{c}}}$)-Eq.} in the blue box) considers a propagating flame front with deviations in flame displacement speed from an unstretched laminar flame, similar to laboratory experiments or classical asymptotic theories~\cite{Clavin_Williams_1982,Matalon82_lam_stretch}. \textcircled{\tiny 2} A micro-scale perspective on the local source term~${\dot{\;\omega_{c}}}$, which is determined by the local mixture state, characterized by the local equivalence ratio~$\phi$, enthalpy~$h$, and radical mass fraction~$Y_{\mathrm{radical}}$. These parameters change according to the local flame structure~($\left|\nabla c \right|$) and geometry~($\kappa$), which leads to a response in~${\dot{\;\omega_{c}}}$ to external perturbations of the flame, e.g. by turbulence. {\color{black}In the limit of ${\mathrm{Le}_k=1}$, $\phi$ and~$h$ remain almost constant across the flame structure, i.e.\ the coupling between~${\dot{\;\omega_{c}}}$ and $\left|\nabla c \right|$ and~$\kappa$ is much weaker than in flames with ${\mathrm{Le}_k\neq1}$.} Note that turbulent fluctuations in the scalar fields ahead of the flame have been omitted.}
\label{fig:PrefDiff_schematic}
%\end{figure*} %%%%%%%%%%%%%%%%%%%%%%%%%%%%%
\end{sidewaysfigure*}

 %%%%%%%%%%%%%%%%%%%%%%%%%%%%%
%%%
To gain a general understanding of the complex processes inside the flame structure of turbulent flames and enable the derivation of accurate models, it is desirable to identify a suitable reduced representation of the multi-parameter system and to establish a connection to canonical laminar reference flames. In the present work, this is pursued by following two analysis pathways according to Fig.~\ref{fig:PrefDiff_schematic}. \par
Starting point is~$\left.\overline{\dot{\;\omega_{c}}}\right|_{\Omega}$ as quantity of interest~(QoI) for engine combustion. In a first step, the flames will be considered from a macroscopic perspective that allows to intuitively demonstrate the impact of differential diffusion on early flame kernel development within the frame of the flame surface density~(FSD) concept~\cite{Bray90_stretch_factor}:
\begin{equation}%%%%%%%%%%%%%%%%%%%%%%%%%%%%% 
\begin{aligned}
%\nonumber
\left.\overline{\dot{\omega}_{c}}\right|_{\Omega} =  
\left(\rho \mathrm{s}_{\mathrm{l,}0}\right) \cdot \overline{\mathrm{I}}_{0} \cdot
 \overline{\Sigma}_{c,\Omega},
 % \left(\overline{\mathrm{I}}_{0,\mathrm{rn}} + \overline{\mathrm{I}}_{0,\kappa} \right)
\label{eq:integr_srcProg}
\end{aligned}
\end{equation} %%%%%%%%%%%%%%%%%%%%%%%%%%%%%
where~$\rho$ is the fluid density, ${\overline{\Sigma}_{c,\Omega}= \overline{A}_{c}/ V_{\Omega}}$~\cite{Boger98_gen_fsd} denotes the global flame surface density and the stretch factor~$\overline{\mathrm{I}}_0$ quantifies deviations of the displacement speed~$\mathrm{s}_{\mathrm{d}}$ from the laminar burning velocity~$\mathrm{s}_{\mathrm{l,}0}$ due to ignition, differential diffusion, and turbulent micro-mixing. To parametrize~$\left.\overline{\dot{\;\omega_{c}}}\right|_{\Omega}$ entirely in terms of flame structure and geometry, $\overline{\mathrm{I}}_0$ can be decomposed into contributions by normal propagation, which is affected by the flame structure, and by tangential diffusion, which represents the flame geometry~\cite{Echekki99_diff_term_split}:
\begin{align} %%%%%%%%%%%%%%%%%%%%%%%%%%%%% 
\nonumber
\overline{\mathrm{I}}_0 & = \frac{\left< \rho {\mathrm{s}}_{\mathrm{d}} \right>_{\mathrm{s,}\Omega}}{\rho\mathrm{s}_{\mathrm{l,}0}}
= \frac{\left< \rho \left( {\mathrm{s}}_{\mathrm{rn}} - D_{\mathrm{th}} \kappa \right) \right>_{\mathrm{s,}\Omega}}{\rho \mathrm{s}_{\mathrm{l,}0}} \\
& \eqdef \overline{\mathrm{I}}_{0,\mathrm{rn}} + \overline{\mathrm{I}}_{0,\kappa}\;,
\label{eq:I0_def_split}
\end{align} %%%%%%%%%%%%%%%%%%%%%%%%%%%%%
where $\left< \cdot \right>_{\mathrm{s}}$ denotes generalized scalar-iso-surface averaging~\cite{Boger98_gen_fsd}, $\mathrm{s}_{\mathrm{rn}}$ is the displacement speed due to reaction and normal diffusion, and $D_{\mathrm{th}}$ is the thermal diffusivity. In this work, the mean curvature~$\kappa$ of iso-surfaces belonging to any scalar field~$\vartheta$ is computed from the normal vector~$n_i$ pointing into the direction of negative scalar gradient, i.e.\ flame kernels have positive global mean curvature:
\begin{align}%%%%%%%%%%%%%%%%%%%%%%%%%%%%% 
%\nonumber
\kappa^{(\vartheta)} &= \frac{\partial n_i^{(\vartheta)}}{\partial x_i\;\;}, \label{eq:def_kappa} \\
n_i^{(\vartheta)} &= \frac{-1}{\left|\nabla \vartheta \right|}\frac{\partial \vartheta}{\partial x_i} . \label{eq:def_nvec}
\end{align} %%%%%%%%%%%%%%%%%%%%%%%%%%%%%
For brevity, the sub/superscript indicating the scalar field will be omitted in most parts below. \par
%For a detailed derivation of global heat release rate parameters, refer to~\cite{Falkenstein19_kernel_Le1_cnf}. \par
%
% ignition at small radius, i.e.\ large positive mean curvature
% curvature and gradient are coupled
%
% ToDo: check systems thinking links! \par
%Systems Thinking
% http://www.systems-thinking.org/intst/int.htm
% http://www.systems-thinking.org/systhink/systhink.htm
% http://www.systems-thinking.org/arch/arch.htm
% http://www.systems-thinking.org/intst/d-3312.pdf
%
The macroscopic perspective based on the reduced flame representation in terms of a propagating front (cf.~\textcircled{\tiny 1} in Fig.~\ref{fig:PrefDiff_schematic}) enables a separate consideration of differential diffusion effects on~$\overline{\mathrm{I}}_{0,\mathrm{rn}}$ (flame structure and the chemical source term) on the one hand, and on~$\overline{\mathrm{I}}_{0,\kappa}$ as well as $\overline{\Sigma}_{c,\Omega}$ (flame geometry) on the other hand. In this way, the actual displacement speed of early flame kernels can be related to a laminar unstretched flame and the transition from a quasi-laminar kernel to a fully developed turbulent flame front may be quantified by the evolution of total flame area~$\overline{A}$, which is related to the curvature distribution of the flame~\cite{Falkenstein19_kernel_Le1_jfm}. While this perspective already captures the external effect of hydrodynamic strain, a differentiated consideration of energy supply by spark ignition, as well as effects of turbulent micro-mixing on the coupled multi-species system inside the flame structure require an additional analysis pathway. \par %, which is meant to identify suitable parameters for relating local turbulent flame behavior to laminar reference flames. \par
Starting point for the micro-scale analysis is the local chemical source term~${\dot{\;\omega_{c}}}$ (cf.~\textcircled{\tiny 2} in Fig.~\ref{fig:PrefDiff_schematic}). In contrast to the displacement speed, the source term only depends on the local state vector [$Y_k,T$], which shall be expressed first by a reduced parameter set in order to characterize differential diffusion effects. As will be shown in Sect.~\ref{sec:loc_hr}, the combination of the local equivalence ratio~$\phi$, local enthalpy~$h$, and the mass fraction of a radical species %with particularly high mobility
is well-suited to capture variations in~${\dot{\;\omega_{c}}}$. While the evolution of each parameter is determined by the respective transport equation, {\color{black}two effects shall be emphasized. First, externally invoked changes in the flame structure~($\left|\nabla c \right|$) will lead to modified transport of thermal and chemical internal energy, hence altering the local mixture state. This effect is intrinsic to both ${\mathrm{Le}>1}$ and ${\mathrm{Le}=1}$ flames and is here represented by the scalar dissipation rate $\chi_c = 2 D_{\mathrm{th}} \left|\nabla c \right|^2$. Note that $\chi_c$ can be considered as the inverse of a characteristic diffusion time scale~\cite{Peters00_book} and may be used to investigate extinction phenomena, e.g. based on a diffusive Damk{\"o}hler number~\cite{Williams00_PECS,Uranakara17_ign_kernel_3d}. \par
%entropy equation in Fig.~\ref{fig:PrefDiff_schematic}, which contains a source term that accounts for irreversible entropy production due to thermal transport~\cite{Masi13_entropy_eq_chi,Borghesi15_entropy_eq_chi}. Note that this term is here rewritten in terms of the scalar dissipation rate $\chi_c = 2 D_{\mathrm{th}} \left|\nabla c \right|^2$ in order to highlight the relation with the thermodynamic (heat) dissipation. Seen from an exergetic point of view, changes in~$\left|\nabla c \right|$ will alter the distribution of combustion irreversibilities attributed to heat conduction, mass diffusion, chemical reaction, and viscous dissipation~\cite{Som08_precs_exergy,Nishida02_entropy_premixed_flame}. \par
The second significant effect on the local mixture state is limited to flames with ${\mathrm{Le}_k\neq1}$ and is represented by }
the (approximate) differential diffusion term~$\mathfrak{D}_{\mathrm{Le}}$ of the enthalpy equation in Fig.~\ref{fig:PrefDiff_schematic}. Although the present $\mathfrak{D}_{\mathrm{Le}}$-formulation may not be valid for high-Karlovitz-number conditions, it highlights the importance of local flame structure ($\left|\nabla c \right|$) and local geometry ($\kappa$) for the occurrence of deviations from laminar flame behavior in case of ${\mathrm{Le}_k\neq1}$. For more details, refer to Part~II of the present study~\cite{Falkenstein19_kernel_Le_II_cnf}. From the \mbox{$\kappa$-} and \mbox{$\left|\nabla c \right|$-}equations given in Fig.~\ref{fig:PrefDiff_schematic}, it is immediately obvious that local flame geometry and structure are coupled, i.e.\ externally invoked changes in $\kappa$ or $\left|\nabla c \right|$ (${\mathrm{s}}_{\mathrm{rn}}$), e.g.\ by hydrodynamic strain, may alter the respective other parameter. Overall, the presence of differential diffusion effects (${\mathrm{Le}_k\neq1}$) leads to a response in heat release rate to externally introduced perturbations in local flame geometry and structure, which may in turn change the $\kappa$ and $\left|\nabla c \right|$ balance. Note that in the limit of ${\mathrm{Le}_k=1}$, the local mixture state will be much less affected {\color{black}by $\kappa$ and $\left|\nabla c \right|$}, {\color{black}i.e.\ a reduced} heat release response {\color{black}to external effects} can be expected. \par
%the local mixture state will be much less affected and heat release response can be expected to be small. \par
While run-to-run variations between different flame kernel realizations computed in the ${\mathrm{Le}_k=1}$ limit were investigated in our previous studies in terms of the \mbox{($\overline{\dot{\;\omega_{c}}}$)-} and \mbox{($\overline{A}$)-Eq.}~\cite{Falkenstein19_kernel_Le1_cnf} as well as the \mbox{($\kappa$)-Eq.}~\cite{Falkenstein19_kernel_Le1_jfm}, the objective of the present work is to isolate the effect of differential diffusion on~$\overline{\dot{\;\omega_{c}}}$ during early flame kernel development. In Part~I of the present study, the global heat release rate is first analyzed from the macroscopic perspective (r.h.s.\ of \mbox{($\overline{\dot{\;\omega_{c}}}$)-Eq.} in Fig.~\ref{fig:PrefDiff_schematic}) to compare the behavior of a flame kernel computed with a realistic engine fuel to previous results computed in the ${\mathrm{Le}_k=1}$ limit. Then, the heat release rate is considered from a micro-scale point of view to identify the parameters that characterize the coupling between~${\dot{\;\omega_{c}}}$ and turbulence/flame interactions in presence of differential diffusion. % by using the optimal estimator technique.
In Part~II of the present study~\cite{Falkenstein19_kernel_Le_II_cnf}, the effect of the local flame structure and geometry on the mixture state ($\phi,h,Y_{\mathrm{H}}$) and source term~${\dot{\;\omega_{c}}}$ is quantitatively investigated by relating the behavior of flame kernels to corresponding planar turbulent flames and laminar reference solutions.
\section{DNS Database}
\label{sec:dns_database}
In order to fundamentally characterize early flame kernel development under SI engine conditions, four flame kernel DNS realizations and one planar flame with ${\mathrm{Le}>1}$  were added to an existing database of three reference flames computed in the ${\mathrm{Le}=1}$ limit (cf.\ {\color{black} Tab.~\ref{tab:dns_datasets}} and Sect.~\ref{sec:turb_dns_database}). Additional comparative analyses are enabled by laminar unstretched and stretched flamelet solutions, which will be summarized in Sect.~\ref{sec:lam_flame_calc}.
\subsection{Three-Dimensional DNS}
\label{sec:turb_dns_database}
The DNS database has been carefully designed to be representative for SI engine part-load conditions. Since the governing equations and numerical methods were extensively described in our previous study on the same flame configurations as in this work, but with all Lewis numbers artificially set to unity~\cite{Falkenstein19_kernel_Le1_cnf,Falkenstein19_kernel_Le1_jfm}, only a brief summary is provided in the following. To enable systematic investigations on differential diffusion effects, four flame kernel realizations and one planar reference flame with realistic (but constant) Lewis numbers have been added to the existing ${\mathrm{Le}=1}$ database. Hence, four flame configurations {\color{black}(kernel/planar, ${\mathrm{Le}=1}$/${\mathrm{Le}>1}$)} are available for this study. While the flame conditions listed in Tab.~\ref{tab:dns_mixture} are fully equivalent to our previous work~\cite{Falkenstein19_kernel_Le1_cnf,Falkenstein19_kernel_Le1_jfm}, some differences in the non-dimensional groups given in Tab.~\ref{tab:dns_params} emerge due to the smaller chemical time scale of the laminar unstretched flame with ${\mathrm{Le}>1}$. In Tab.~\ref{tab:dns_mixture}, the mixture thermodynamic state is given by the pressure~$p^{\left(0\right)} $ and temperature of the unburned gas~$T_{\mathrm{u}}$ with equivalence ratio~$\phi_{\mathrm{u}}$. 
\begin{table}%[!htb] %%%%%%%%%%%%%%%%%%%%%%%%%%%%%
\caption{Flame conditions in the DNS.}
\vspace{0.1cm}
\centering
%\begin{tabular}{P{.22\textwidth}|P{.22\textwidth}}
\begin{tabular}{P{.35\textwidth}|P{.35\textwidth}} %% review_12pt
\hline
~ & ~ \\ [-10pt]
Property & Value  \\
~ & ~ \\ [-10pt]
\hline
~ & ~ \\ [-10pt]
Mixture   & Iso-Octane/Air \\ %[3pt]
$p^{\left(0\right)} $ & $6$ bar  \\
$T_{\mathrm{u}}$ & $600$ K \\
$\phi_{\mathrm{u}}$ & $1.0$ (homogeneous)\\
$\mathrm{s}_{\mathrm{l}}^{\sm{0}}$ & $\boldsymbol{0.73} \; |\;${\color{black}$0.63$} m/s \\
$l_{\mathrm{f}}$ & $\boldsymbol{69.1}  \; |\;${\color{black}$71.3$} $\mu \mathrm{m}$ \\
%sL_u=0.7303045
%lF=6.91092E-05
Flow Field & Decaying h.i.t. \\
Combust. Regime & Thin Rct. Zones\\
\hline
~ & ~ \\ [-10pt]
$\mathrm{Le}_{\mathrm{eff}}$ & $\boldsymbol{2.0} \; |\; 1.0$ \\
\hline
\end{tabular}
\label{tab:dns_mixture}
\end{table} %%%%%%%%%%%%%%%%%%%%%%%%%%%%%
 Note that the turbulent flow field corresponds to decaying homogeneous isotropic turbulence, which explains the DNS parameter ranges given in Tab.~\ref{tab:dns_params}. The turbulent integral length scale and eddy turnover time are denoted by~$l_{\mathrm{t}}$ and~$\tau_{\mathrm{t}}$, $\eta$ is the Kolmogorov length scale, while the laminar flame thickness and initial flame diameter are referred to as~$l_{\mathrm{f}}$ and~$D_{0}$, respectively.
% (Le1-Le2)/Le2=0.238
Due to computational restrictions, the wavenumber range covered by the DNS is smaller than in an actual engine. Hence, the integral length scale in the simulations is two to three times smaller than in reality, while the Karlovitz number is closely matched to actual engine conditions {\color{black}at part load~\cite{Linse09_engine_regime_diagr}}. For an illustration of the DNS conditions in the regime diagram of turbulent combustion {and a discussion of the engine relevance of the DNS configuration}, refer to~\cite{Falkenstein19_kernel_Le1_cnf}. Although the Karlovitz number in that study was slightly higher than in the present work due to the ${\mathrm{Le}=1}$ simplification, it was shown that no thickening of the averaged flame structure occurs. \par
\begin{table}%[!htb] %%%%%%%%%%%%%%%%%%%%%%%%%%%%%
\caption{Engine \citep{Heywood94_COMODIA,Heim11_engine_turb} and DNS characteristic numbers ($\mathrm{Le}>1$).}
\vspace{0.1cm}
\centering
%\begin{tabular}{P{.138\textwidth}|P{.138\textwidth}|P{.138\textwidth}}
\begin{tabular}{P{.23\textwidth}|P{.23\textwidth}|P{.23\textwidth}}  % review_12pt
\hline
~ & ~ & ~ \\ [-10pt]
Parameter & Engine & DNS ($t_{\mathrm{init}} - t_{\mathrm{end}}$)  \\
~ & ~ & ~ \\ [-10pt]
\hline
~ & ~ & ~ \\ [-10pt]
$\mathrm{Re}_{\mathrm{t}} $ & $100-2390$  & $ 385 - 222$\\[3pt]
$\frac{u_{\mathrm{rms}}}{\mathrm{s}_{\mathrm{l}}^{\sm{0}}}$& $2-15$ & $ 5.9 - 2.8 $\\[3pt]
$\mathrm{Ka} $& $1-6$ & $10.6 - 3.2$\\[3pt]
$\mathrm{Da}$ &$1-100$ & $ 1.9 - 4.6$\\[3pt]
%$\frac{l_{\mathrm{f}}}{\eta}$& & & $8.0 - 4.4$\\[2pt] % TODO: needed? -> Ka
$\frac{l_{\mathrm{t}}}{\eta}$& 100 -200 & $87.2 - 57.2$\\[3pt]  % TODO: check
$\frac{l_{\mathrm{t}}}{l_{\mathrm{f}}}$& $20 - 147$ & $10.9 - 13.0$\\[3pt]
\hline
~ & ~ & ~ \\ [-10pt]
$\frac{D_{0}}{l_{\mathrm{t}}}$ &$<1.0$ & $ 0.3 \; |\; \infty $\\[3pt]
\hline
\end{tabular}
\label{tab:dns_params}
\end{table} %%%%%%%%%%%%%%%%%%%%%%%%%%%%%
In order to quantitatively assess the impact of differential diffusion on the heat release rate, a detailed chemical reaction scheme is required~\cite{Hilbert04_PRECS_detailed_chem}. As in our previous study, a modified kinetic model based on the skeletal iso-octane mechanism by Pitsch and Peters~\cite{Pitsch96_iso_octane_mech} was used. Validation results for the reaction scheme calibrated to the present DNS conditions are provided as supplementary material. {\color{black}Diffusive scalar transport was solved based on the Curtiss-Hirschfelder approximation~\cite{Hirschfelder54_diff_model} with additional consideration of the Soret effect (thermodiffusion). With respect to the accuracy of the diffusion model employed here, refer to the work by van~Oijen et al.~\cite{vanOijen16_fgm_precs}, where premixed $\mathrm{CH}_4$/$\mathrm{H}_2$ flame solutions computed with the Curtiss-Hirschfelder approximation and with a complex diffusion model based on the Stefan-Maxwell equations were compared.} \par %The calibrated mechanism predicts very similar results to the CaltechMech~\cite{Blanquart15_caltech_mech}, as shown in the supplementary material for a spherical expanding flame (ToDo).\par
A brief summary of the DNS setup is given in Tab~\ref{tab:dns_setup}. {\color{black} All flames were initialized in the same realization of a decaying homogeneous isotropic turbulent flow field, which was generated from a combination of multiple forced turbulence solutions as described in our previous study~\cite{Falkenstein19_kernel_Le1_cnf}.} One important difference between the flame kernel datasets and the planar reference flames is the initialization method. Flame kernels were ignited by a source term in the temperature equation, which results in an early growth phase comparable to engine experiments reported in literature (cf.~\cite{Falkenstein19_kernel_Le1_jfm}). By contrast, a laminar unstretched flame was imposed into the turbulent flow field as initial condition for the planar-flame DNS to avoid strong dilatation due to the larger burned volume. {\color{black} Multiple flame kernel realizations were computed by initializing each simulation with the same flow field, while varying the ignition location by more than five integral length scales between simulation runs. To provide flame statistics for the particularly critical early flame kernel development phase and to limit the overall computational costs, one reference flame kernel computed for a physical time of ${t_{\mathrm{sim.}}=3.4\cdot\tau_{\mathrm{t}}}$ has been supplemented by three realizations that were simulated until $t_{\mathrm{sim.}}=1.0\cdot\tau_{\mathrm{t}}$ (cf. Tab.~\ref{tab:dns_setup}). The planar flame simulation was stopped after $t_{\mathrm{sim.}}=2.8\cdot\tau_{\mathrm{t}}$, which is approximately the time when the flame front can be considered as fully developed based on near-zero total flame area rate-of-change. While flame kernels were computed in an isochoric setting with periodic boundary conditions, an outlet boundary condition was used to achieve isobaric conditions for the planar flame. For the main flame kernel reference case, the pressure increase at ${t_{\mathrm{sim.}}=3.4\cdot\tau_{\mathrm{t}}}$ was below 8.5\,\%.} \par
%Nevertheless, the dataset is well-suited as a canonical reference case, as will be shown below. \par
%
\begin{table}%[!htb] %%%%%%%%%%%%%%%%%%%%%%%%%%%%%
\caption{Computational setup of the DNS.}
\vspace{0.1cm}
\centering
%\begin{tabular}{P{.22\textwidth}|P{.22\textwidth}}
\begin{tabular}{P{.35\textwidth}|P{.35\textwidth}} %% review_12pt
\hline
~ & ~ \\ [-10pt]
Property & Value  \\
~ & ~ \\ [-10pt]
\hline
~ & ~ \\ [-10pt]
Grid Size   & $960^3$ \\ %[3pt]
 ~ & ~ \\ [-10pt]
Domain Size & $15 \cdot l_{\mathrm{t}}$ \\ %[3pt]
 ~ & ~ \\ [-10pt]
Navier-Stokes Eq. & Low-Mach-Approx.~\cite{Mueller98_lowMach} \\
 ~ & ~ \\ [-10pt]
Transport Model & Curt.-H.~\cite{Hirschfelder54_diff_model}, const.-Le  \\
 ~ & ~ \\ [-10pt]
Soret Effect & yes \\
 ~ & ~ \\ [-10pt]
Chem. Mechanism &  26 Spec., based on~\cite{Pitsch96_iso_octane_mech} \\
 ~ & ~ \\ [-10pt]
\multicolumn{1}{l|}{\makebox[20pt][l]{}\color{black}\emph{Flame Kernels:}}  & ~ \\
\multicolumn{1}{l|}{\makebox[20pt][l]{}\tabitem \color{black}Initialization} & \multicolumn{1}{l}{\makebox[10pt][l]{}\tabitem\color{black}Ign. Heat Source}\\
\multicolumn{1}{l|}{\makebox[20pt][l]{}\tabitem \color{black}Boundry. Cond.} & \multicolumn{1}{l}{\makebox[10pt][l]{}\tabitem\color{black}$x,y,z$-dir.: periodic}\\
\multicolumn{1}{l|}{\makebox[20pt][l]{}\tabitem \color{black}Sim. Time} & \multicolumn{1}{l}{\makebox[10pt][l]{}\tabitem\color{black}1x ($t_{\mathrm{sim.}}=3.4\cdot\tau_{\mathrm{t}}$),} \\
\multicolumn{1}{l|}{} & \multicolumn{1}{l}{\makebox[10pt][l]{}\tabitem\color{black}3x ($t_{\mathrm{sim.}}=1.0\cdot\tau_{\mathrm{t}}$)\,} \\
 ~ & ~ \\ [-10pt]
\multicolumn{1}{l|}{\makebox[20pt][l]{}\color{black}\emph{Planar Flame:}}  & ~ \\
\multicolumn{1}{l|}{\makebox[20pt][l]{}\tabitem \color{black}Initialization} & \multicolumn{1}{l}{\makebox[10pt][l]{}\tabitem\color{black}Lam. Flamelet}\\
\multicolumn{1}{l|}{\makebox[20pt][l]{}\tabitem \color{black}Boundry. Cond.} & \multicolumn{1}{l}{\makebox[10pt][l]{}\tabitem\color{black}$x$-dir.: symm./outlet,}\\
\multicolumn{1}{l|}{} & \multicolumn{1}{l}{\makebox[10pt][l]{}\tabitem\color{black}$y,z$-dir.: periodic}\\
\multicolumn{1}{l|}{\makebox[20pt][l]{}\tabitem \color{black}Sim. Time} & \multicolumn{1}{l}{\makebox[10pt][l]{}\tabitem\color{black}($t_{\mathrm{sim.}}=2.8\cdot\tau_{\mathrm{t}}$)} \\
\hline
\end{tabular}
\label{tab:dns_setup}
\end{table} %%%%%%%%%%%%%%%%%%%%%%%%%%%%%
To simplify the analysis and make the dataset more accessible for modelling, a new reaction progress variable~$\zeta$ has been defined by the solution of the transport equation~\cite{Falkenstein19_kernel_Le1_cnf}
\begin{equation}
\frac{\partial \left( \rho \zeta \right)}{\partial t} + \frac{\partial \left( \rho u_j \zeta \right)}{\partial x_j} =
	\frac{\partial}{\partial x_j} \left( \rho D_{\mathrm{th}} \frac{\partial \zeta }{\partial x_j} \right) + \dot{\;\omega_{\zeta}}.
\label{eq:c0_eq}
\end{equation}
The chemical source term was chosen as the sum of source terms of main product species:
\begin{equation}
\dot{\;\omega_{\zeta}} = \dot{\omega}_{\mathrm{H}_2}+\dot{\omega}_{\mathrm{H}_2\mathrm{O}} + \dot{\omega}_{\mathrm{CO}} + \dot{\omega}_{\mathrm{CO}_2}.
\label{eq:c0_eq_src_term}
\end{equation}
Solving~Eq.~(\ref{eq:c0_eq}) in the DNS simplifies the analysis of the overall normal and tangential diffusion effects, as compared to a progress variable definition that results in a balance equation with more complex diffusive transport terms. {\color{black}Differences between the r.h.s.\ of~Eq.~(\ref{eq:c0_eq}) and the balance equation of a progress variable $c = Y_{\mathrm{H}_2} + Y_{\mathrm{H}_2\mathrm{O}} + Y_{\mathrm{CO}} + Y_{\mathrm{CO}_2}$ are quantitatively discussed in Sect.~3 in the supplementary material.} The decomposition of the first term in the r.h.s of~Eq.~(\ref{eq:c0_eq}) according to Echekki and Chen~\cite{Echekki99_diff_term_split} will be employed in Sect.~\ref{sec:integr_hr} to analyze the decomposed stretch factor (cf.\ Eq.~(\ref{eq:I0_def_split})).
\subsection{Laminar Flame Calculations}
\label{sec:lam_flame_calc}
To relate local conditions inside the turbulent flame structures to laminar flamelet solutions, several one-dimensional flames were computed with the \textit{FlameMaster} code~\cite{Pitsch98_FM}. Steady, unstretched laminar flames~(sub/superscript `lam') were solved based on the algorithm by Smooke et al.~\cite{Smooke83_unstr_sL}. To generate stretched reference flames, the premixed back-to-back counterflow configuration~(sub/superscript `cff') has been selected in analogy to previous modeling studies~\cite{Knudsen13_strained_model,Trisjono16_strained_model_test}. The numerical solutions were obtained based on the similarity coordinate formulation assuming a potential flow in the far field~\cite{DixonLewis85_counterflow_similarity}. For reaction mechanism validation and assessment of early flame kernel growth, laminar spherical flames~(sub/superscript `sph')  have been computed in Lagrangian coordinates~\cite{Maas88_1d_sph_flame} using DASSL~\cite{Petzold82_dassl}. \par
A detailed analyis of the laminar unstretched flame structure in iso-octane/air flames in terms of a very similar kinetic scheme as used in the present study was provided by Pitsch and Peters~\cite{Pitsch96_iso_octane_mech} and is not repeated here. For a discussion of differential diffusion effects on flame structure and extinction in the premixed counterflow configuration, refer to the modeling study by van~Oijen and de Goey ~\cite{vanOijen02_premixedCntFlow_Le} or the review paper by Lipatnikov and Chomiak~\cite{Lipatnikov05_PRECS}. \par
\section{Results}
\label{sec:results}
As described in Fig.~\ref{fig:PrefDiff_schematic}, the analysis of heat-release-rate response has been conducted according to two logical pathways. In Sect.~\ref{sec:integr_hr}, differential diffusion effects on the \textit{global heat release rate} will be discussed from the macroscopic perspective by considering two equivalent flame kernel datasets computed with ${\mathrm{Le}>1}$ and ${\mathrm{Le}=1}$, respectively. For the micro-scale analysis, the ability of several reduced parameter sets to capture differential-diffusion-induced fluctuations in the \textit{local heat release rate} of planar flames will be quantified in Sect.~\ref{sec:loc_hr}. The main results will be used to explain the Lewis number effects on the global heat release rate of flame kernels observed in Fig.~\ref{fig:integr_srcProg}.
%According to Fig.~\ref{fig:PrefDiff_schematic}, heat-release-rate variations are linked to spurious fluctuations in enthalpy and local equivalence ratio, which will be discussed in Sect.~\ref{sec:loc_enth_phi} and related to the local flame geometry and flame structure, parametrized by curvature and scalar gradient magnitude. Finally, the impact of characteristic flame kernel/turbulence interactions~\cite{Falkenstein19_kernel_Le1_jfm} in ${\mathrm{Le}>1}$ flames will be explored in Sect.~\ref{sec:pref_diff_curv_distr}.
%
%-------------------------------------------------------------------------------------------------------------------------------------------
\subsection{Global Heat Release Rate Evolution}
\label{sec:integr_hr}
To better understand the strong impact of differential diffusion on~$\left.\overline{\dot{\omega}_{c}}\right|_{\Omega}$ that was shown in Fig.~\ref{fig:integr_srcProg}, the stretch factor~$\overline{\mathrm{I}}_0$ and the flame surface density~$\overline{\Sigma}_{c,\Omega}$ will be considered according to the r.h.s.\ of the \mbox{($\overline{\dot{\;\omega_{c}}}$)-Eq.} in Fig.~\ref{fig:PrefDiff_schematic}. In this way, differential diffusion effects on the local burning velocity can be distinguished from effects on flame area evolution.
Note that the results in Fig.~\ref{fig:integr_srcProg} were plotted as function of time, which is relevant for the engine application. However, this representation implies that differences in the integrated source term accumulate due to growing differences in flame kernel size. In order to rigorously analyze the role of differential diffusion during flame development, it is desirable to enable comparisons of data extracted from flames with similar geometry and size. Hence, results will be shown as function of kernel radius hereafter. The radius{\color{black}~$R_{50}$} is here defined as the median of the distance distribution, measured between all flame points and the geometric center of the flame. {\color{black}It should be noted that at the same flame kernel radius, the flames are exposed to slightly different turbulence intensity since the turbulence decays over time. At a flame radius of one integral length scale $R_{50}=l_{\mathrm{t}}$ (cf.\ Fig.~\ref{fig:integr_srcProg}), the turbulent kinetic energy differs by~7\,\%. } \par
\begin{figure}
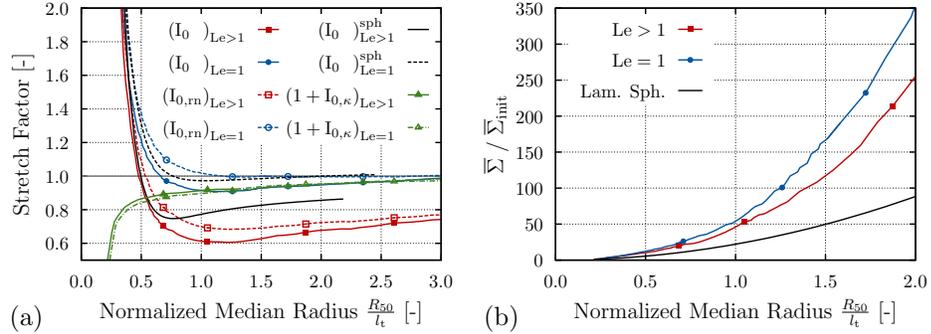
 %%%%%%%%%%%%%%%%%%%%%%%%%%%%%
\centering
\begin{minipage}[b]{0.45\textwidth}
  \graphicspath{{./data/FLAME_KERNEL_DNS_01/181003_curv_strain_kin_restor_diss_divg_skew/}}
  \centering
  % trim={<left> <lower> <right> <upper>}
  % \hfill\vspace{0.1cm}
\input{template/change_font_10.tex} % review_12pt
  \makebox[0pt][l]{\quad(a)}\scalebox{0.7}{\input{./data/FLAME_KERNEL_DNS_01/181003_curv_strain_kin_restor_diss_divg_skew/kernel_dns_mean_I0_Sd_1d_size_01_02_JFM_HALF.tex}}
\input{template/change_font_12.tex} % review_12pt
% Plot_I0_Sd_1d_curv_size_01_02_ltx_JFM.gp
\end{minipage}
\begin{minipage}[b]{0.45\textwidth}
  \graphicspath{{./data/FLAME_KERNEL_DNS_01/181003_curv_strain_kin_restor_diss_divg_skew/}}
  \centering
%trim={<left> <lower> <right> <upper>}
%\hfill\vspace{0.1cm}
  %\makebox[0pt][l]{\quad(a)}\includegraphics[trim={10cm -4cm 40cm -6cm},clip,width=5.5cm]{}
\input{template/change_font_10.tex} % review_12pt
    \makebox[0pt][l]{\quad(b)}\scalebox{0.7}{\input{./data/FLAME_KERNEL_DNS_01/181003_curv_strain_kin_restor_diss_divg_skew/flame_kernel_gen_fsd_size_K01_K02_JFM_HALF.tex}}
\input{template/change_font_12.tex} % review_12pt
\end{minipage}
\caption{Flame Kernel: Mean stretch factor~(a) and domain-integrated generalized flame surface density~(b) as function of kernel radius. Superscript `sph' corresponds to a laminar spherical expanding flame.}
\label{fig:kernel_I0_Sigma}
\end{figure}

A comparison of the stretch factors evaluated for both the ${\mathrm{Le}>1}$ and the ${\mathrm{Le}=1}$ turbulent flame kernel is provided in Fig.~\ref{fig:kernel_I0_Sigma}\,(a). Initially, $\overline{\mathrm{I}}_0$ is dominated by the effect of spark ignition, which affects the early flame kernels until a radius of approximately one integral length scale is reached. This behavior is similar to actual engines, as discussed in our previous study~\cite{Falkenstein19_kernel_Le1_jfm}. At the kernel size of one integral length scale, differential diffusion effects reduce~ $\overline{\mathrm{I}}_0$ by 33\,\%. 
%. i.e.\ the reduction in integral heat release rate is initially dominated by the burning velocity. 
%
% t0_startIgn3 = 1.71523e-04
% l0_startIgn3 = 7.34370e-04
% urms_startIgn3 = 4.28147e+00
%
% LE > 1
% interp_val_col.sh ciao_time_size.dat 2 7.3437e-4 1
%  1.561265e-04  = t(r=l_t)
% interp_val_col.sh DNS_FM_ref_data_SELECT_t.txt 1 1.561265e-04 4
%  7.433681e-01 = sl
% interp_val_col.sh DNS_FM_ref_data_SELECT_t.txt 1 1.561265e-04 5
%  3.643047 = rho
% interp_val_col.sh ciao_all_dat_files_disp_speed_RHO_cond_curv_gen_surf_avg.dat 1 1.561265e-04 2
%  1.658924  = rho*sd  = I0*rho*SL => I0 = 0.613 -> 33% smaller w.r.t. Le=1
%  0.455 = sd -> 21% smaller w.r.t. Le=1
% interp_val_col.sh ciao_all_dat_files_disp_speed_RHO_cond_curv_gen_surf_avg.dat 1 1.561265e-04 5
%  1.892044 = rho*s_rn = I0_rn*rho*SL
%  0.5194 = s_rn -> 19% smaller w.r.t. Le=1
%%%%%%%%%
% LE = 1
% interp_val_col.sh ciao_time_size.dat 2 7.3437e-4 1
%  1.316511e-04  = t(r=l_t)
% interp_val_col.sh DNS_FM_ref_data_Le1_SELECT_t.txt 1 1.316511e-04 4
%  6.274280e-01 = sl
% interp_val_col.sh DNS_FM_ref_data_Le1_SELECT_t.txt 1 1.316511e-04 5
%  3.643192e+00 = rho
% interp_val_col.sh ciao_all_dat_files_disp_speed_RHO_cond_curv_gen_surf_avg.dat 1  1.316511e-04 2
%  2.091269  = rho*sd = I0*rho*SL => I0 = 0.915
%  0.574 = sd 
% interp_val_col.sh ciao_all_dat_files_disp_speed_RHO_cond_curv_gen_surf_avg.dat 1  1.316511e-04 5
%  2.329084 = rho*s_rn  = I0_rn*rho*SL
%  0.6394
This is a stronger reduction than observed for two laminar spherical expanding flames, which were ignited in the same way as the DNS and are plotted in Fig.~\ref{fig:kernel_I0_Sigma}\,(a) for reference~(superscript `sph'). As shown in Part~II of the present study~\cite{Falkenstein19_kernel_Le_II_cnf}, turbulence detrimentally alters the flame structure by hydrodynamic strain, in addition to the unfavorable effect of large positive curvature on the heat release rate of ${\mathrm{Le}>1}$ flame kernels. It should be noted that~$\overline{\mathrm{I}}_0$ quantifies the deviation of the actual flame displacement speed from the respective laminar burning velocity, which is in fact lower in case of ${\mathrm{Le}=1}$ ($\mathrm{s}_{\mathrm{l,Le=1}}^{\sm{0}} / \mathrm{s}_{\mathrm{l,Le>1}}^{\sm{0}} = 0.84$). At the kernel size of one integral length scale, the net displacement speed of the ${\mathrm{Le}>1}$ flame is~21\,\% lower than in the ${\mathrm{Le}=1}$ limit. {\color{black}
This is interesting, since the unstretched laminar burning velocity for the ${\mathrm{Le}>1}$  case is actually larger than for the ${\mathrm{Le}=1}$ case. Similar findings have also been reported by Savard and Blanquart~\cite{Savard15_Le_effects_C7H16_highKa} who studied n-heptane/air flames both for ${\mathrm{Le}>1}$ and ${\mathrm{Le}=1}$.
%The fact that the turbulent ${\mathrm{Le}>1}$ flame kernel exhibits lower displacement speed than in the ${\mathrm{Le}=1}$ limit although the opposite is true under laminar flow conditions is in line with previous results for lean n-heptane/air flames with $\mathrm{Le}>1$ and ${\mathrm{Le}=1}$ studied by Savard and Blanquart~\cite{Savard15_Le_effects_C7H16_highKa}.
} \par
{\color{black}To gain additional insights, the stretch factor is here decomposed} into a component representative of flame normal propagation and a curvature effect (cf.\ Eq.~(\ref{eq:I0_def_split})). 
Obviously, the normal-propagation stretch factor $\overline{\mathrm{I}}_{0,\mathrm{rn}}$ is significantly reduced in the ${\mathrm{Le}>1}$ flame after ignition effects have decayed. According to the \mbox{($\mathrm{s}_{\mathrm{rn}}$)-Eq.} in Fig.~\ref{fig:PrefDiff_schematic}, this can be due to changes in the chemical source term~$\dot{\;\omega_c}$ and the scalar gradient magnitude $\left|\nabla c \right|$, which is related to the local flame thickness.
From the definition of $\overline{\mathrm{I}}_{0,\kappa}$ (cf.\ Eq.~(\ref{eq:I0_def_split})), it can be expected that differential diffusion only has a minor effect on this term, since the correlation between curvature and the diffusion coefficient is rather weak and mean curvature is not altered by changes in Lewis number. This is confirmed by the results shown in Fig.~\ref{fig:kernel_I0_Sigma}\,(a). Note that $\overline{\mathrm{I}}_{0,\kappa}$ is a measure for diffusive transport of~$\zeta$ in iso-surface-tangential directions, while curvature effects on the chemical source term are contained in~$\overline{\mathrm{I}}_{0,\mathrm{rn}}$. \par
%In the ${\mathrm{Le}=1}$ case, $\overline{\mathrm{I}}_{0,\mathrm{rn}}$ approaches unity, i.e.\ the flame displacement speed in the DNS is equal to the laminar burning velocity of an unstretched flame. However, the reaction/normal-diffusion balance of the flame kernel is affected by the tangential diffusion term due to mean flame curvature for a long time, which is balanced by normal diffusion(ToDo: maybe add supp.\ material).\par
%%
The domain-integrated generalized flame surface density $\overline{\Sigma}_{\zeta,\Omega}$ is plotted in Fig.~\ref{fig:kernel_I0_Sigma}\,(b) as function of kernel radius. Obviously, the realistic engine fuel leads to a suppression of net flame area production due to thermal-diffusive effects, which is in agreement with planar flame results reported in literature~\cite{Haworth92_dns,Trouve94_fsd,Bell07_Le_effect_wrinkling}. 
%Since both flame kernels develop inside the same flow field, it can be expected that differences in flame area growth are primarily caused by altered flame displacement speed and/or curvature distribution (cf.~\cite{Falkenstein19_kernel_Le1_jfm} for a detailed analysis of flame area dynamics in the ${\mathrm{Le}=1}$ flame). ToDo: check if statement is needed
When the flame kernels have reached a radius of one integral length scale, which corresponds to the time when ignition effects have mostly decayed~\cite{Falkenstein19_kernel_Le1_jfm}, the flame surface density of the ${\mathrm{Le}>1}$ flame is reduced by 15\,\% compared to the ${\mathrm{Le}=1}$ case. 
%
% LE > 1
% interp_val_col.sh ciao_time_size.dat 2 7.3437e-4 1
%  1.561265e-04  = t(r=l_t)
% interp_val_col.sh ciao_all_dat_files_disp_speed_gen_surf_avg.dat 1 1.561265e-04 6
%  1.849706 = fsd(non-normalized)
% interp_val_col.sh mon_io_PROG0_min_max_t_mid_SELECT_t.dat 1 1.561265e-04 3
%  2.520450e-01 = c0
% sigm0_Le2=0.159916
% fsd / c0 / sigm0_Le2 = 45.89
%%%%%%%%%
% LE = 1
% interp_val_col.sh ciao_time_size.dat 2 7.3437e-4 1
%  1.316511e-04  = t(r=l_t)
% interp_val_col.sh ciao_all_dat_files_disp_speed_gen_surf_avg.dat 1 1.316511e-04 6
%  2.179951 = fsd(non-normalized)
% interp_val_col.sh mon_io_PROG0_min_max_t_mid_SELECT_t.dat 1 1.316511e-04 3
%  2.588086e-01 = c0
% sigm0_Le1=0.156374
% fsd / c0 / sigm0_Le1 = 53.86
%
%This effect can be attributed to the normal-propagation term in the flame area balance equation, as will be shown below. 
Note that both turbulent flames exhibit significantly larger surface areas than a perfect sphere with the same radius, as indicated in the figure. From the preceding discussion it can be concluded that the reduction in integral heat release rate attributed to differential diffusion effects is initially dominated by the burning velocity, while noticeable differences in flame area exist as well. \par
\begin{figure}
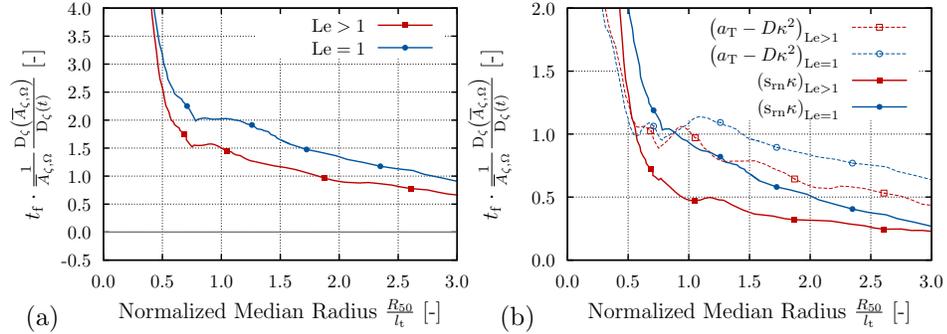
 %%%%%%%%%%%%%%%%%%%%%%%%%%%%%
\centering
\begin{minipage}[b]{0.45\textwidth}
  \graphicspath{{./data/FLAME_KERNEL_DNS_01/181003_curv_strain_kin_restor_diss_divg_skew/}}
  \centering
%trim={<left> <lower> <right> <upper>}
%\hfill\vspace{0.1cm}
%  \makebox[0pt][l]{\quad(a)}\includegraphics[trim={0cm -0.75cm 0cm 0cm},clip,width=7cm]{./data/FLAME_PLANAR_DNS_01/190316_local_equiv_ratio_curv/t_3.600E-04/}
\input{template/change_font_10.tex} % review_12pt
  \makebox[0pt][l]{\quad(a)}\scalebox{0.7}{\input{./data/FLAME_KERNEL_DNS_01/181003_curv_strain_kin_restor_diss_divg_skew/kernel_dns_mean_1ovA_dAdt_size_K01_K02_JFM_HALF.tex}}
\input{template/change_font_12.tex} % review_12pt
\end{minipage}
\begin{minipage}[b]{0.45\textwidth}
  \graphicspath{{./data/FLAME_KERNEL_DNS_01/181003_curv_strain_kin_restor_diss_divg_skew/}}
  \centering
  % trim={<left> <lower> <right> <upper>}
  % \hfill\vspace{0.1cm}
\input{template/change_font_10.tex} % review_12pt
    \makebox[0pt][l]{\quad(b)}\scalebox{0.7}{\input{./data/FLAME_KERNEL_DNS_01/181003_curv_strain_kin_restor_diss_divg_skew/kernel_dns_mean_1ovA_dAdt_PROPAGATION_sumStrDiss_size_K01_K02_JFM_HALF.tex}}
\input{template/change_font_12.tex} % review_12pt
\end{minipage}
\caption{Flame kernel area rate-of-change as function of kernel radius: Net effect~(a), contribution by normal propagation shown separately~(b).}
\label{fig:kernel_planar_dAdt_vs_time}
\end{figure}

 %%%%%%%%%%%%%%%%%%%%%%%%%%%%%
%
%In the next section it will be shown that the initial reduction in $\overline{\Sigma}_{\zeta,\Omega}$ can be entirely attributed to the normal-propagation term in the flame area balance equation, which in turn depends on the flame displacement speed.
%-------------------------------------------------------------------------------------------------------------------------------------------
%\subsection{Differential Diffusion Effects on Flame Kernel Area Growth}
%\label{sec:pref_diff_area}
In order to further investigate how differential diffusion reduces the overall flame surface area (or~$\overline{\Sigma}_{\zeta,\Omega}$) and consequently the integral heat release rate of the ${\mathrm{Le}>1}$ flame kernel (cf.\ Fig.~\ref{fig:kernel_I0_Sigma}\,(b)), the total flame area rate-of-change is considered~\cite{Falkenstein19_kernel_Le1_cnf}:
\begin{equation}
\frac{1}{\overline{A}_{\zeta,\Omega}} \frac{\mathrm{D}_{\zeta} \left( \overline{A}_{\zeta,\Omega} \right)}{\mathrm{D}_{\zeta}\left(t\right)} = 
\frac{\int_{\Omega}\left(\left(\mathrm{s}_{\mathrm{rn}} \kappa \right) - D_{\mathrm{th}} \kappa^2 + a_{\mathrm{t}}\right)\left|\nabla \zeta\right|\diff V}{\int_{\Omega}\left|\nabla \zeta\right|\diff V}.
\label{eq:one_ov_A_dAdt_split}
\end{equation}
Results extracted from the ${\mathrm{Le}>1}$ and the ${\mathrm{Le}=1}$ datasets are compared as function of kernel size in Fig.~\ref{fig:kernel_planar_dAdt_vs_time}. First, we focus on the evolution of the l.h.s.\ of Eq.~(\ref{eq:one_ov_A_dAdt_split}) to assess differences in net area rate-of-change. As shown in Fig.~\ref{fig:kernel_planar_dAdt_vs_time}\,(a), flame area production is overall positive due to the presence of mean flame curvature, but generally higher in the ${\mathrm{Le}=1}$ case. At ($R_{50}=1.0\cdot l_{\mathrm{t}}$), when ignition effects have mostly decayed, differential diffusion leads to a reduction in flame area production rate by 25\,\%. From the strong reduction in flame normal-propagation velocity observed in Fig.~\ref{fig:kernel_I0_Sigma}\,(a), it can be expected that the first term in the r.h.s.\ of Eq.~(\ref{eq:one_ov_A_dAdt_split}), i.e.\ the change in flame area due to normal propagation of a curved surface, plays a major role in causing this difference. Hence, this term is plotted separately from the tangential strain and scalar dissipation terms in Fig.~\ref{fig:kernel_planar_dAdt_vs_time}\,(b). In this way, it becomes obvious that the normal-propagation term is responsible for almost all differences in early flame area growth  in the range (${R_{50}<1.0\cdot l_{\mathrm{t}}}$). In particular, the sum of the remaining two terms is initially very similar for both Lewis number cases, while differential diffusion reduces the area production through normal propagation by up to 50\,\%. During this early phase, the effect of differential diffusion on flame kernel area growth may be similar to laminar flames (cf.\ Fig.~\ref{fig:kernel_I0_Sigma}\,(a)). As the flame kernels grow in size, differences in the curvature distributions develop (cf.\ Part~II~\cite{Falkenstein19_kernel_Le_II_cnf}), which cause differences in heat release rate that feed back into the tangential strain term in Eq.~(\ref{eq:one_ov_A_dAdt_split}). \par

%According to Eq.~(\ref{eq:integr_srcProg}), the integral progress variable source term $\left.\overline{\dot{\omega}_{\zeta}}\right|_{\Omega}$ is determined by the parameters normalized mean burning velocity (stretch factor) $ \overline{\mathrm{I}}_0$ and flame surface area per volume $\overline{\Sigma}_{\zeta,\Omega}$. 
%
 %%%%%%%%%%%%%%%%%%%%%%%%%%%%%
%\textcolor{black}{Since the reduction in global heat release rate due to differential diffusion has been shown to be governed by~${\mathrm{s}}_{\mathrm{rn}}$ through the direct effect on~$\overline{\mathrm{I}}_{0,\mathrm{rn}}$ as well as the change in~$\overline{\Sigma}_{\zeta,\Omega}$, the local heat release rate will be analyzed in more detail in the following section.}
The preceding discussion has shown that differential diffusion has a strong impact on the burning rate during flame kernel development under engine-relevant conditions especially because of its influence on flame normal propagation. Both flame area and flame displacement speed are significantly lower in the flame computed with a realistic transportation fuel, i.e.\ with ${\mathrm{Le}>1}$. Compared to an unstretched laminar flame, the net displacement speed is reduced by up to 40\,\% due to differential and tangential diffusion.
% The combined effect of reduction in normal-propagation velocity (Lewis-number-dependent) and negative contribution by tangential diffusion (Lewis-number-independent) (ToDo: depends on PROG definition) decreases the net displacement speed by up to 40\,\% compared to an unstretched laminar flame.
As this minimum in displacement speed occurs at very small flame radii, 
%(when flame acceleration by ignition effects diminishes), 
the overall flame kernel behavior may still be very sensitive to stochastic interactions with the turbulent flow field~\cite{Falkenstein19_kernel_Le1_jfm}. Hence, the discussed burning-rate evolution is expected to be particularly critical for the occurrence of~CCV in SI engines. \par
%-------------------------------------------------------------------------------------------------------------------------------------------
\subsection{Local Heat Release Rate Variations}
\label{sec:loc_hr}
To further investigate the strong Lewis-number dependence of the mean normal-propagation stretch factor~$\overline{\mathrm{I}}_{0,\mathrm{rn}}$ (cf.\ Sect.~\ref{sec:integr_hr}), the parameter interactions inside the flame structure that govern the local heat release rate will be analyzed in the following. 
%Here, the objective is to identify a reduced representation of the coupling between turbulent perturbations of the flame and the local heat release rate (cf.\ Fig.~\ref{fig:PrefDiff_schematic}) based on quantitative criteria. 
Since the physical and chemical phenomena inside an unsteady flame structure are already quite complex, the following analysis is simplified by limiting ourselves to the fully developed, statistically planar-turbulent-flame datasets. {\color{black}Most importantly, the planar flame configuration provides sufficient statistics for a rigorous comparison between the ${\mathrm{Le}>1}$ and the ${\mathrm{Le}=1}$ cases.} %Transient flame kernel development will be considered in Sects.~\ref{sec:kernel_loc_enth_phi} and~\ref{sec:pref_diff_area}, after a fundamental understanding of the role of differential diffusion under the present engine-relevant conditions has been established and the turbulent-flame results have been related to canonical laminar flame configurations. 
%This approach corresponds to a systematic backward analysis according to Fig.~\ref{fig:PrefDiff_schematic}.
%\textcolor{red}{It will be shown that differential diffusion effects in engine-relevant fuel/air mixtures cause a strong negative correlation between curvature and heat release rate, which is disadvantageous for flame kernels with initially high, only slowly decaying positive mean curvature. Even in flame regions with zero curvature, the heat release rate is detrimentally affected by hydrodynamic strain, similarly to a laminar counterflow configuration. }
It will be shown that in addition to a reduction in mean heat release rate, tremendous fluctuations in heat release rate exist throughout the flame structure of the ${\mathrm{Le}>1}$ flame. To parametrize the heat release rate in ${\mathrm{Le}\neq1}$ flames, the combination of local equivalence ratio, enthalpy, and radical availability will be quantitatively compared to other parameter choices and confirmed as suitable parameter set, in agreement with individual findings from literature that were used to design Fig.~\ref{fig:PrefDiff_schematic}. \par
\begin{figure} %%%%%%%%%%%%%%%%%%%%%%%%%%%%%
\centering
\begin{minipage}[b]{0.45\textwidth}
  \graphicspath{{./data/FLAME_PLANAR_DNS_01/181229_local_equiv_ratio_avg/pics/}}
  \centering
%trim={<left> <lower> <right> <upper>}
%\hfill\vspace{0.1cm}
\input{template/change_font_10.tex} % review_12pt
    \makebox[0pt][l]{\quad(a)}\includegraphics[trim={35cm -6cm 40cm -2cm},clip,width=7cm]{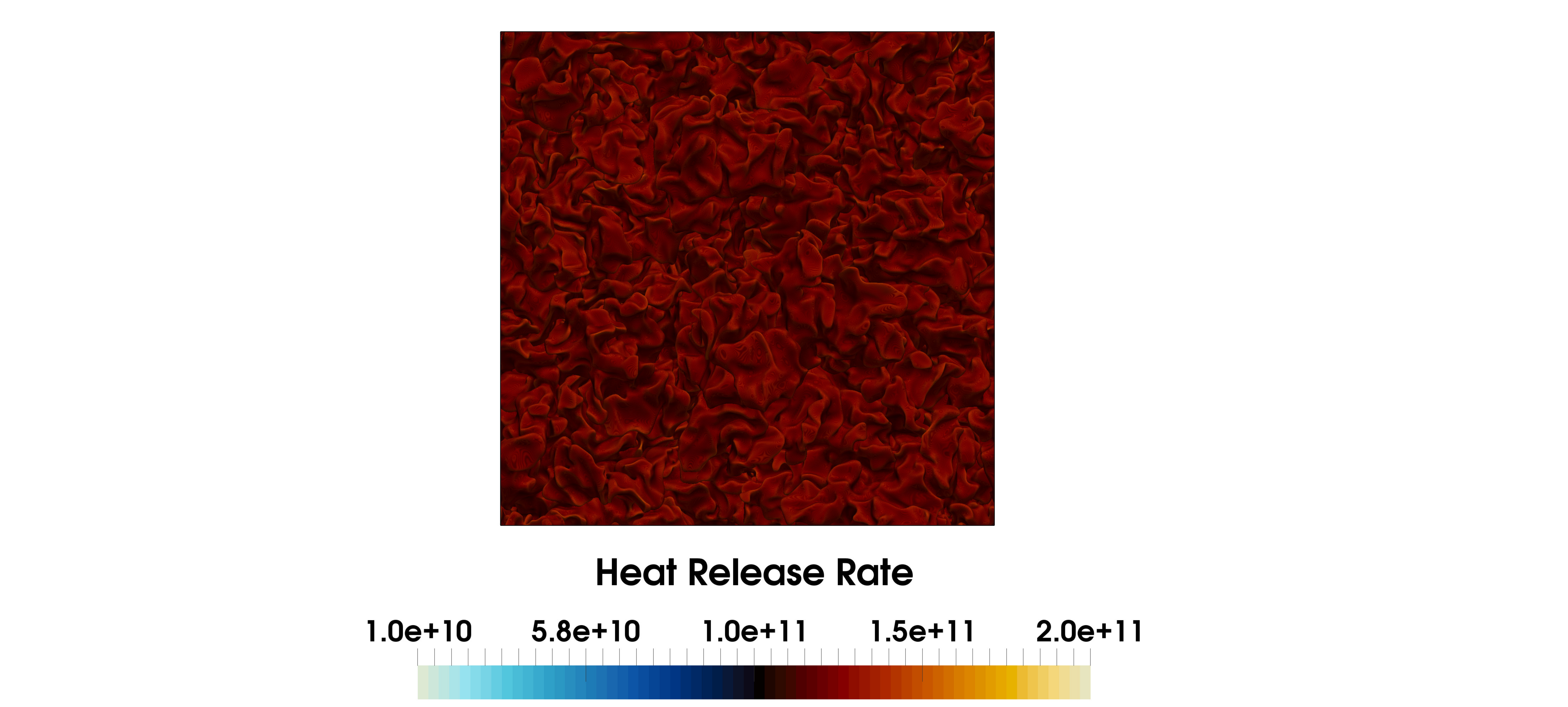} 
\input{template/change_font_12.tex} % review_12pt
\end{minipage}
\begin{minipage}[b]{0.45\textwidth}
  \graphicspath{{./data/FLAME_PLANAR_DNS_02/181229_local_equiv_ratio_avg/pics/}}
  \centering
  % trim={<left> <lower> <right> <upper>}
  % \hfill\vspace{0.1cm}
\input{template/change_font_10.tex} % review_12pt
    \makebox[0pt][l]{\quad(b)}\includegraphics[trim={35cm -6cm 40cm -2cm},clip,width=7cm]{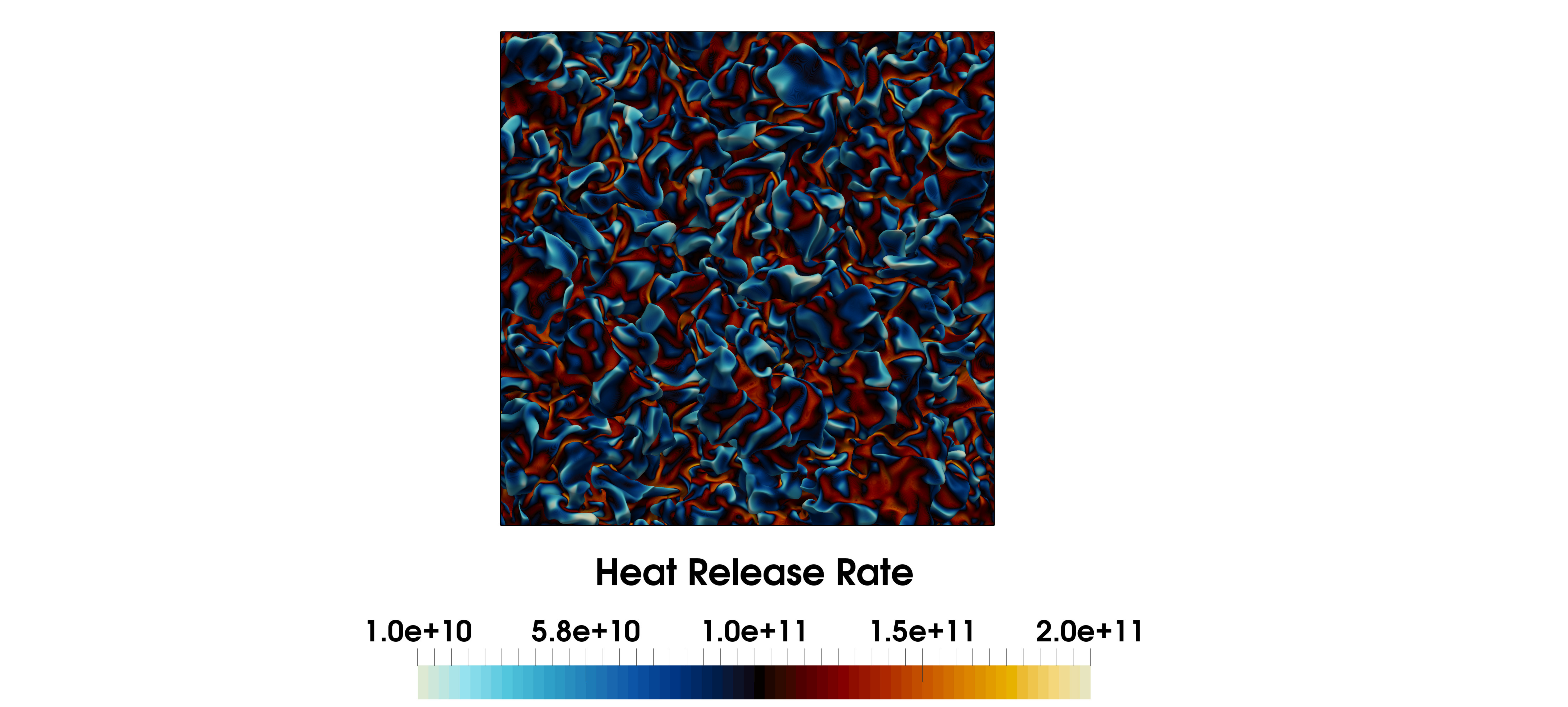}
\input{template/change_font_12.tex} % review_12pt
\end{minipage}
\caption{Planar Flame: $T_{\mathrm{maxHR}}$ iso-surface colored by local heat release rate for the ${\mathrm{Le}=1}$ dataset~(a) and the engine-relevant ${\mathrm{Le}>1}$ flame~(b) at ${t=2.8\,\tau_{\mathrm{t}}}$.}
\label{fig:planar_hr_vis}
\end{figure}

To get a first impression of the impact of differential diffusion on the spatial heat release distribution, the temperature iso-surfaces corresponding to maximum heat release rate in the laminar unstretched flame (`$T_{\mathrm{maxHR}}$') have been extracted from both planar-turbulent-flame datasets at ${t=2.8\,\tau_{\mathrm{t}}}$ and are colored by heat release rate in Fig.~\ref{fig:planar_hr_vis}. In the ${\mathrm{Le}=1}$ dataset, the heat release rate distribution is almost uniform. {\color{black}This already suggests that in the ${\mathrm{Le}=1}$ limit, the coupling of $\chi_c$ and~${\dot{\;\omega_{c}}}$ (cf.\ Fig.~\ref{fig:PrefDiff_schematic}) does not lead to significant spatial variations in~${\dot{\;\omega_{c}}}$ under the present flow conditions.} Conversely, the more engine-relevant flame with ${\mathrm{Le}>1}$ features similarly high source term magnitudes only in regions of negative curvature, while large parts of the plotted iso-surface exhibit significantly reduced heat release rates. This observation is consistent with the correlation between the fuel consumption rate and curvature that was previously shown for an n-heptane/air flame in the broken reaction zones regime~\cite{Savard15_Le_effects_C7H16_highKa} and the correlation between the consumption speed and curvature in ${\mathrm{Le}\neq1}$ flames located in the corrugated flamelet regime~\cite{Haworth92_dns}. Correlations between the local heat release rate and flame structure/geometry are discussed in Part~II of the present study~\cite{Falkenstein19_kernel_Le_II_cnf}. For a visualization of differential diffusion effects  during flame kernel development, refer to \mbox{Figs.~S-6} and~\mbox{S-7} of the supplementary material. \par
\begin{figure}
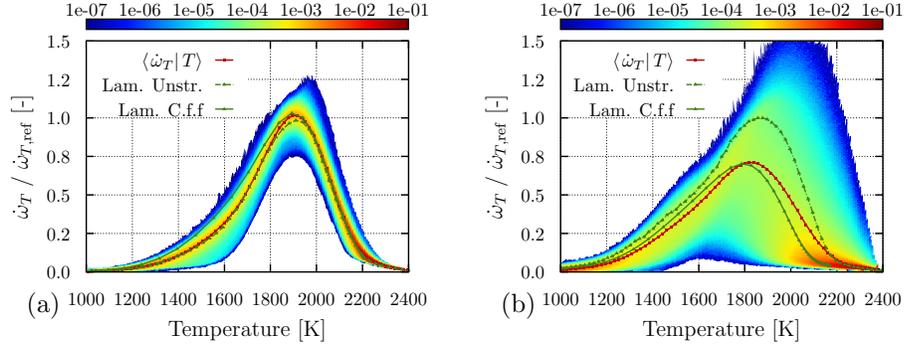
 %%%%%%%%%%%%%%%%%%%%%%%%%%%%%
\centering
\begin{minipage}[b]{0.45\textwidth}
  \graphicspath{{./data/FLAME_PLANAR_DNS_02/190504_HR_condT_optEst/t_3.6E-04//}}
  \centering
%trim={<left> <lower> <right> <upper>}
%\hfill\vspace{0.1cm}
  %\makebox[0pt][l]{\quad(a)}\includegraphics[trim={10cm -4cm 40cm -6cm},clip,width=5.5cm]{}
\input{template/change_font_10.tex} % review_12pt
  \makebox[0pt][l]{\quad(a)}\scalebox{0.7}{\input{./data/FLAME_PLANAR_DNS_02/190504_HR_condT_optEst/t_3.6E-04/planar_dns_t3p6e-4_jPDF_HR_vs_T_P02_ltx_JFM_HALF.tex}}
\input{template/change_font_12.tex} % review_12pt
\end{minipage}
\begin{minipage}[b]{0.45\textwidth}
  \graphicspath{{./data/FLAME_PLANAR_DNS_01/190504_HR_condT_optEst/t_3.6E-04//}}
  \centering
  % trim={<left> <lower> <right> <upper>}
  %\hfill\vspace{0.1cm}
\vspace{0.1cm}
\input{template/change_font_10.tex} % review_12pt
  \makebox[0pt][l]{\quad(b)}\scalebox{0.7}{\input{./data/FLAME_PLANAR_DNS_01/190504_HR_condT_optEst/t_3.6E-04/planar_dns_t3p6e-4_jPDF_HR_vs_T_P01_ltx_JFM_HALF.tex}}
% /home/tf236914/FLAME_KERNEL/FLAME_PLANAR_DNS_01/190504_HR_condT_optEst/t_3.6E-04
% Plot_jPDF_HR_vs_T_P01_ltx_JFM.gp
%  jPDF_hr_temp_gp.dat
%  README_jPDF_ltx.txt
% ../../190316_G_local_PHI_condT/t_3.6E-04/cond_surf_avg_gradT_HR_vs_T.dat
%  main_only_plots_condMean_HR_pdf.m
% ../../laminar_flames/CIAO/190321_flame_struct/cond_surf_avg_gradT_HR_vs_T.dat
\input{template/change_font_12.tex} % review_12pt
\end{minipage}
\vspace{0.1cm}
\caption{Planar Flame: Joint-PDFs of heat release rate and temperature for the ${\mathrm{Le}=1}$ dataset~(a) and the engine-relevant ${\mathrm{Le}>1}$ flame~(b) at ${t=2.8\,\tau_{\mathrm{t}}}$. {\color{black}Similar results are shown for the flame kernel configuration in Fig.~\mbox{S-5} in the supplementary material} }
\label{fig:planar_jPDF_hr_vs_temp}
\end{figure}

 %%%%%%%%%%%%%%%%%%%%%%%%%%%%%
To enable a more quantitative understanding of differential diffusion effects on the heat release rate distribution across the flame structure, joint-PDFs of heat release rate and temperature are shown in Fig.~\ref{fig:planar_jPDF_hr_vs_temp}. Additionally, conditional mean results are plotted for comparison with laminar reference data. In the ${\mathrm{Le}=1}$ flame, the heat release rate of an unstretched premixed flame is almost perfectly recovered (cf.\ Fig.~\ref{fig:planar_jPDF_hr_vs_temp}\,(a)). In the ${\mathrm{Le}>1}$ case, a different behavior is expected due to the reduction in stretch factor that was observed in the corresponding flame kernel dataset (cf.\ Fig.~\ref{fig:kernel_I0_Sigma}\,(a)). As shown in Fig.~\ref{fig:planar_jPDF_hr_vs_temp}\,(b), the maximum conditional mean heat release rate of the turbulent planar flame is reduced by 29\;\% with respect to the laminar unstretched value. {\color{black}A very similar reduction was observed by Savard et al.~\cite{Savard17_si_engine_dns} for a slightly lean iso-octane/air flame with comparable Karlovitz number, but computed at a higher pressure and temperature of 20\,bar and 800\,K.} 
% FLAME_PLANAR_DNS_01/190504_HR_condT_optEst/t_3.6E-04
% print_min_max_col.py ../../190316_G_local_PHI_condT/t_3.6E-04/cond_surf_avg_gradT_HR_vs_T.dat
%  1.17260e+11
% print_min_max_col.py ../../laminar_flames/CIAO/190321_flame_struct/cond_surf_avg_gradT_HR_vs_T.dat
%  1.65040e+11
For reference, data extracted from a back-to-back counterflow flame (c.f.f.) solution has been added to the plots in Fig.~\ref{fig:planar_jPDF_hr_vs_temp}. 
The strained laminar flame solutions have been selected based on the condition  ${\left|\nabla T \right|_{{\mathrm{maxHR}},\mathrm{cff}} = \left< \left. \left|\nabla T \right|\; \right|\; \left(T=T_{\mathrm{maxHR}}, \kappa=0\right)\right>}$ and will be analyzed in more detail in Part~II~\cite{Falkenstein19_kernel_Le_II_cnf}. {\color{black}Note that in the ${\mathrm{Le}=1}$ limit (cf.\ Fig.~\ref{fig:planar_jPDF_hr_vs_temp}\,(a)), the effect of~$\chi_c$ on the mixture state sketched in Fig.~\ref{fig:PrefDiff_schematic} seems to be small under the present flow conditions. The ${\mathrm{Le}>1}$ flame additionally exhibits a heat release rate dependence on curvature, which also explains } 
the difference between the turbulent and the laminar profiles on the high-temperature side of the peak in Fig.~\ref{fig:planar_jPDF_hr_vs_temp}\,(b). \par% can be reduced by conditioning the data on zero curvature, which is not shown for clarity. \par
%(ToDo counterflow: laminar flame: mass balance requires variable mass flux as fct. of strain. 1. stretch changes gradients, which triggers differential diffusion 2. strain changes residence time, which reduces temperature and leads to extinction even in $\mathrm{Le}<1$ case 3. Radiative heat losses increase at low stretch rates (thicker flame), but is not considered here. cf Law1989, WilliamsPRECS00) \par
Besides the reduction in mean heat release rate observed in the planar, ${\mathrm{Le}>1}$ flame as compared to the corresponding ${\mathrm{Le}=1}$ dataset, visual inspection of Fig.~\ref{fig:planar_jPDF_hr_vs_temp} immediately suggests a strong impact of differential diffusion on variations in local heat release rate. While the data points in the ${\mathrm{Le}=1}$ case shown in Fig.~\ref{fig:planar_jPDF_hr_vs_temp}\,(a) are distributed in a rather narrow band around the conditional mean profile, tremendous scatter from almost extinguished flamelets to levels well above the unstretched reference flame is present in the ${\mathrm{Le}>1}$ dataset. 
Note that the occurrence of flame elements with particularly low heat release rates was previously identified as a distinct feature of the thin reaction zones regime~\cite{Shim13_h2_dns_hrr_fluct_thin_rct_zones}. {\color{black}Here, temperature rather than the progress variable~$\zeta$ was chosen as independent variable for this analysis to facilitate comparisons with other studies~\cite{Savard15_Le_effects_C7H16_highKa,Lapointe15_appendix_avg,Savard17_si_engine_dns}. Joint-PDFs of $\dot{\;\omega_{\zeta}}$ and~$\zeta$ are provided in Fig.~\mbox{S-4} in the supplementary material.}  {\color{black}The preceding discussion on the planar flame datasets is additionally supported by qualitatively similar results for the flame kernel configuration provided in Fig.~\mbox{S-5} in the supplementary material.} \par
%This is consistent with the results by Shim et al.~\cite{Shim13_h2_dns_hrr_fluct_thin_rct_zones}, who identified the occurrence of flame elements with particularly low heat release rates as a distinct feature of the thin reaction zones regime.  \par
 %%%%%%%%%%%%%%%%%%%%%%%%%%%%%
As a next step, we seek to identify the parameters that govern the local heat-release-rate response to differential diffusion effects, i.e.\ the observed \textit{high conditional variance} and \textit{reduced conditional mean}. The findings will be used to relate the detrimental reduction in global heat release rate due to ${\mathrm{Le}>1}$ (cf.\ Fig.~\ref{fig:integr_srcProg}) to the mixture conditions inside the flame structure. To this end, we make use of the optimal estimator analysis technique~\cite{Moreau06_opt_est,Trisjono15_dns_analysis_review}. The primary objective of the following parameter tests is not to derive or check a specific model, but to quantify the ability of a selected input parameter set to predict local heat release rate throughout the flame structure, irrespective of a functional model expression. In this regard, the optimal estimator concept is a powerful tool which is based on two main ideas. First, the conditional mean of a dataset is considered as an ideal reference model, i.e.\ the optimal estimator. Second, the expected error of a model based on the selected input parameters~$\Pi$ cannot be smaller than the standard deviation from the respective optimal estimator, which is cast into an error measure called the irreducible error:
\begin{equation}%%%%%%%%%%%%%%%%%%%%%%%%%%%%% 
\begin{aligned}
%\nonumber
\epsilon^2 \left(\,\Pi\, \right) = \left\Vert \left. \left[\dot{\omega}_T - 
\left\Vert\left.\dot{\omega}_T\right|\Pi\, \right\Vert\; \right]^2 \,\right| \Pi\, \right\Vert.
% \frac{1}{N} \sum_{i=1}^N
\label{eq:irred_error}
\end{aligned}
\end{equation} %%%%%%%%%%%%%%%%%%%%%%%%%%%%%
To reduce fitting errors which may pollute the optimal estimator results, artificial neural networks have been employed as suggested by Berger et al.~\cite{Berger18_opt_est}. Since we are specifically interested in the role of differential diffusion, all irreducible errors have been normalized by $\epsilon^2_{\mathrm{Le}=1}\left(T\right)$, i.e.\ the error value computed for the ${\mathrm{Le}=1}$ flame when using temperature as the only model parameter. Hence, a normalized error value of unity would imply that the scatter around the conditional mean yields the same error sum as computed from the data points plotted in Fig.~\ref{fig:planar_jPDF_hr_vs_temp}\,(a). The expected irreducible errors associated to heat release rate predictions from different input parameter sets in the ${\mathrm{Le}>1}$ flame are listed in Tab.~\ref{tab:irr_err}. When only temperature is used to parametrize heat release rate (cf.\ Fig.~\ref{fig:planar_jPDF_hr_vs_temp}\,(b)), the irreducible error is 5.5~times larger than in the ${\mathrm{Le}=1}$ flame. \par
The empirical input parameter selection closely follows the schematic shown in Fig.~\ref{fig:PrefDiff_schematic} in a backward-analysis fashion, starting from the local heat release rate~$\dot{\omega}_T$ as quantity of interest. 
%Obviously, the presence of~$\mathfrak{D}_{\mathrm{Le}}$ that couples the flame surface geometry and flame structure parameters~$\kappa$ and~$\left|\nabla T \right|$ with the local mixture state characterized by ($h,\phi,Y_{\mathrm{radical}}$), leads to a high variance in heat release rate (for the present conditions inside the thin reaction zones regime). This coupling is drastically weakened in case of  ${\mathrm{Le}_k=1}$, i.e.\ ${\mathfrak{D}_{\mathrm{Le}}=0}$. \par
From a chemical kinetic point of view it might be interesting to begin the backward analysis with the identification of pathways that differ between the ${\mathrm{Le}>1}$ and the  ${\mathrm{Le}=1}$ flames. Similar analyses have been published for different fuels~\cite{Dasgupta17_H2_pathways_turb,Dasgupta18_C12H26_pathways_turb}, including systematic variations of the Lewis number~\cite{Lapointe15_appendix_avg}, but are not pursued here. Still, finding suitable marker species which contain most of the observed heat release rate variations may serve as a starting point for future work and may suggest control variables for flamelet modeling. Due to the direct influence of individual species mass fractions on chemical source terms, these are here referred to as \textit{first-level parameters} (reaction/diffusion-controlled). Among common marker species used for modeling of hydrocarbon flames~\cite{Knudsen13_strained_model}, molecular hydrogen was found to yield an irreducible error $\epsilon^2\left(T,Y_{\mathrm{H_2}}\right)$ comparable to the reference error $\epsilon^2_{\mathrm{Le}=1}\left(T\right)$. {\color{black}The selection of suitable marker species candidates will be further discussed below.} \par
% (ToDo Joint PDF of HR and some marker species, cf.~\cite{Dinesh17_spherical_H2_dns_equiv_ratio} Figs. 11 \& 12). 
As a next step, heat release rate has been parametrized by the stoichiometry and energy content of the local mixture. The chosen \textit{second-level parameters} (diffusion-controlled) for a reduced representation of the mixture state are local equivalence ratio~\cite{deGoey99_stretch_theory,JPope99_local_equiv_ratio_premixed_flame} and enthalpy~\cite{Ashurst87_enth_Le}, which are not changed by chemical source terms. {\color{black}The local equivalence ratio~$\phi$ is defined by
\begin{equation}%%%%%%%%%%%%%%%%%%%%%%%%%%%%% 
\phi = \frac{2\;\mathfrak{X}_{\mathrm{C}} + 0.5\; \mathfrak{X}_{\mathrm{H}}}{\mathfrak{X}_{\mathrm{O}}}; \,\,\,\,\,\mathfrak{X}_{\ell} = \sum_{k=1}^{N_{\mathrm{sp}}} a_{k,\ell}\, X_k ,
\end{equation}%%%%%%%%%%%%%%%%%%%%%%%%%%%%% 
where $\mathfrak{X}_{\ell}$ denotes the element mole fraction of element~$\ell$, $X_k$ is the mole fraction of species~$k$, and~$a_{k,\ell}$ is the number of atoms of element~$\ell$ contained in a molecule of species~$k$~\cite{JPope99_local_equiv_ratio_premixed_flame}.} 
Recall that the unburned mixture is homogeneous, i.e.\ variations in both parameters can only occur due to differential and thermodiffusion. Although enthalpy and equivalence ratio are not independent quantities~\cite{vanOijen02_premixedCntFlow_Le}, $\epsilon^2\left(T, \phi, h\right)$ is clearly beneficial compared to using only
%is the smallest normalized error value of all ${\mathrm{Le}>1}$ tests, which is not achieved by
one of both parameters. Note that the irreducible error $\epsilon^2\left(T, h\right)$ is larger than $\epsilon^2\left(T, \phi\right)$, which seems in contradiction to the early work by Ashurst et al.~\cite{Ashurst87_enth_Le}, who identified enthalpy as a suitable parameter to characterize flame dynamics in ${\mathrm{Le}\neq1}$ mixtures. This discrepancy to the present results is likely a feature of the multi-species system as opposed to single-step-chemistry DNS. The irreducible error can be further reduced by supplementing the second-level parameter set by one radical mass fraction~\cite{Echekki96_radical_differential_diff,Chen00_H2_dns,Aung02_exp_turb_flame_pref_diff}. In particular, $\epsilon^2\left(T, \phi, h,Y_{\mathrm{H}}\right)$ yields the smallest normalized error value of all ${\mathrm{Le}>1}$ tests. In fact, this parameter set gives the same errors in both the ${\mathrm{Le}>1}$ and  ${\mathrm{Le}=1}$ flames, which is in agreement with the combination of individual findings reported in literature (cf.\ Fig.~\ref{fig:PrefDiff_schematic}). Although molecular hydrogen was identified as the most suitable first-level-parameter, the combination $\left(T, \phi, h,Y_{\mathrm{H}_2}\right)$ is less effective than considering radical availability, which {\color{black}is due to a strong correlation between~$Y_{\mathrm{H}_2}$ and~$\phi$ as indicated by the almost identical irreducible errors for $\left(T,Y_{\mathrm{H}_2}\right)$ and $\left(T,\phi\right)$. This observation is supported by the joint-PDF of~$\phi$ and~$Y_{\mathrm{H}_2}$ for a given temperature and~$Y_{\mathrm{H}}$ as shown in Fig.~\ref{fig:planar_jpdf_opt_est_spec}\,(a)}. \par 
As enthalpy and equivalence ratio variations are linked to flame curvature and scalar gradient magnitudes through~$\mathfrak{D}_{\mathrm{Le}}$ (cf.\ Fig.~\ref{fig:PrefDiff_schematic}), $\kappa$ and~$\left|\nabla T \right|$ form the \textit{third-level parameters} (kinematic) for a reduced description of the flame geometry and structure. For simplicity, only the temperature field has been considered to evaluate both parameters, which is a reasonable approximation in the present datasets (cf.\ Part~II~\cite{Falkenstein19_kernel_Le_II_cnf}). %(cf.~\ref{sec:pref_diff_enth}). 
However, the approximate expression for $\mathfrak{D}_{\mathrm{Le}}$ given in Fig.~\ref{fig:PrefDiff_schematic} is certainly inexact and will result in an increased irreducible error. While accounting for curvature in $\epsilon^2\left(T, \kappa \right)$ leads to some reduction in the expected irreducible error, the dominant third-level parameter seems to be $\left|\nabla T \right|$. 
% ToDo: maybe refer to the planar-flame illustration where this result might be counter-intuitive
{\color{black}Note that~$\left|\nabla T \right|$ may affect the mixture state through~$\mathfrak{D}_{\mathrm{Le}}$ and~$\chi$ as sketched in Fig.~\ref{fig:PrefDiff_schematic}.} 
Although~$\left|\nabla T \right|$ and~$\kappa$ are not independent~\cite{Chakraborty05_Le_effect_curv}, the parameter combination yields a lower error $\epsilon^2\left(T, \kappa, \left|\nabla T\right| \right)$, which is still more than 50\,\% higher than the reference error $\epsilon^2_{\mathrm{Le}=1}\left(T\right)$. It should be noted that such deterioration in performance of an ideal model is expected as the relation between the quantity of interest (heat release rate) and input parameters becomes weaker. This increased variance compared to the ${\mathrm{Le}=1}$ flame might be due to local changes in species mass fractions and gradients induced by turbulent micro-mixing (i.e.\ the temperature field may be less representative) or finite response time of the second-level (diffusion-controlled) parameters~($\phi,h$) to changes in curvature and scalar gradients. The latter effect is investigated by testing a \textit{fourth-level parameter set} (kinematic/dynamic), which contains an additional quantity representative for the time rate-of-change of the third-level parameters. Here, the tangential strain rate has been selected as a measure for hydrodynamic changes to the flame structure (cf.\ \mbox{($\left|\nabla T\right|$)-Eq.} in Fig.~\ref{fig:PrefDiff_schematic}) and geometry. In fact, a correlation between local enthalpy and strain rate was already shown to exist in the earliest DNS with ${\mathrm{Le}\neq1}$~\cite{Ashurst87_enth_Le}. Since it is well-known that curvature and strain are correlated~\cite{Haworth92_dns,Renou98_exp_strain_curv_corr}, it is not surprising that the irreducible errors $\epsilon^2\left(T, \left|\nabla T\right|, a_{\mathrm{t}} \right)$ and $\epsilon^2\left(T, \left|\nabla T\right|, \kappa \right)$ are identical. Still, accounting for both curvature and tangential strain in addition to~$\left|\nabla T\right|$ is even more effective. An almost identical irreducible error is achieved by replacing the tangential strain rate by the gradient rate-of-change (l.h.s.\ of \mbox{($\left|\nabla T\right|$)-Eq.} in Fig.~\ref{fig:PrefDiff_schematic}) to include the flame structure dynamics into $\epsilon^2\left(T, \kappa, \left|\nabla T\right|, \frac{1}{\left|\nabla T\right|} \, \frac{\mathrm{D}_{T}\left( \left|\nabla T\right| \right) }{\mathrm{D}_{T} \left( t \right) } \right)$.  \par
\begin{table}%[!htb] %%%%%%%%%%%%%%%%%%%%%%%%%%%%%
\caption{Irreducible error intrinsic to heat release rate predictions from different input parameter sets (normalized by the temperature-based error value computed from the ${\mathrm{Le}=1}$ flame). {\color{black}Data in parentheses corresponds to the ${\mathrm{Le}=1}$ flame.}}
\vspace{0.1cm}
\centering
%\begin{tabular}{P{.22\textwidth}|P{.22\textwidth}}
\begin{tabular}{P{.35\textwidth}|P{.35\textwidth}} %% review_12pt
\hline
%\\ [-10pt] % review_12pt
Parameter Set $\Pi$ & $\epsilon^2_{\mathrm{Le}>1}\left(\Pi\right) \;/\; \epsilon^2_{\mathrm{Le}=1}\left(T\right) $  \\[-2pt] % review_12pt dist was added
%\\ [-10pt] % review_12pt
\hline
$T$ &  \multicolumn{1}{r}{$5.49$\quad($1.00$)\makebox[16pt][l]{}} \\[-2pt] % review_12pt dist was added
\hline
%$T,Y_{\mathrm{H}}$ & $1.84$ \\[-2pt] % review_12pt dist was added
$T,Y_{\mathrm{H}}$ & \multicolumn{1}{r}{$1.84$\quad\color{black}($0.59$)\makebox[16pt][l]{}} \\[-2pt] % review_12pt dist was added
%$T,Y_{\mathrm{H}_2}$ & $0.93$ \\[-2pt] % review_12pt dist was added
%
$T,Y_{\mathrm{H}_2}$ & \multicolumn{1}{r}{$0.93$\quad\color{black}($0.53$)\makebox[16pt][l]{}} \\[-2pt] % review_12pt dist was added
${\color{black}T,Y_{\mathrm{H}_2},Y_{\mathrm{H}}}$ & \multicolumn{1}{r}{${\color{black}0.54}$\quad\color{black}($0.35$)\makebox[16pt][l]{}} \\[-2pt] % review_12pt dist was added
%${\color{black}\left(\mathrm{Le}=1:\; T,Y_{\mathrm{H}_2} \,|\, T,Y_{\mathrm{H}} \right)}$ & ${\color{black}\left( 0.53 \,|\, 0.59\right)}$ \\[-2pt] % review_12pt dist was added
%${\color{black}\left(\mathrm{Le}=1:\; T,Y_{\mathrm{H}_2},Y_{\mathrm{H}}\right)}$ & ${\color{black}\left( 0.35\right)}$ \\[-2pt] % review_12pt dist was added
\hline
$T,h$ & $1.88$ \\[-2pt] % review_12pt dist was added
$T,h,Y_{\mathrm{H}}$ & $1.07$ \\[-2pt] % review_12pt dist was added
$T,h,Y_{\mathrm{H}_2}$ & $0.77$ \\[-2pt] % review_12pt dist was added
%$T,\phi$ & $0.94$ \\[-2pt] % review_12pt dist was added
%
$T,\phi$ &  \multicolumn{1}{r}{$0.94$\quad\color{black}($0.54$)\makebox[16pt][l]{}} \\[-2pt] % review_12pt dist was added
$T,\phi,Y_{\mathrm{H}}$ & $0.50$ \\[-2pt] % review_12pt dist was added
$T,\phi,Y_{\mathrm{H}_2}$ & $0.76$ \\[-2pt] % review_12pt dist was added
$T,\phi,h$ & $0.64$ \\[-2pt] % review_12pt dist was added
$T,\phi,h,Y_{\mathrm{H}_2}$ & $0.55$ \\[-2pt] % review_12pt dist was added
%$\boldsymbol{T,\phi,h,Y_{\mathrm{H}}}$ &  $\boldsymbol{0.30}$ \\[-2pt] % review_12pt dist was added
$\boldsymbol{T,\phi,h,Y_{\mathrm{H}}}$ &  \multicolumn{1}{r}{$\boldsymbol{0.30}$\quad($\boldsymbol{0.31}$)\makebox[11pt][l]{}} \\[-2pt] % review_12pt dist was added 
%${\color{black}\left(\mathrm{Le}=1:\; T,\phi\right)}$ & ${\color{black}\left( 0.54\right)}$ \\[-2pt] % review_12pt dist was added
%$\boldsymbol{\left(\mathrm{Le}=1:\; T,\phi,h,Y_{\mathrm{H}}\right)}$ & $\boldsymbol{\left( 0.31\right)}$ \\[-2pt] % review_12pt dist was added
\hline
$T,\kappa$ & \multicolumn{1}{r}{$3.84$\quad\color{black}($0.69$)\makebox[16pt][l]{}} \\[-2pt] % review_12pt dist was added
%$T,\left|\nabla T\right|$ & $1.78$ \\[-2pt] % review_12pt dist was added
%
$T,\left|\nabla T\right|$ & \multicolumn{1}{r}{$1.78$\quad\color{black}($0.74$)\makebox[16pt][l]{}} \\[-2pt] % review_12pt dist was added
$T,\left|\nabla T\right|,\kappa$ & $1.52$ \\[-2pt] % review_12pt dist was added
%${\color{black}\left(\mathrm{Le}=1:\; T,\left|\nabla T\right|\right)}$ & ${\color{black}\left( 0.74\right)}$ \\[-2pt] % review_12pt dist was added
\hline
$T,a_{\mathrm{t}}$ & \multicolumn{1}{r}{$4.68$\quad\color{black}($0.93$)\makebox[16pt][l]{}} \\[-2pt] % review_12pt dist was added
$T,\kappa,a_{\mathrm{t}}$ & \multicolumn{1}{r}{$3.46$\quad\color{black}($0.59$)\makebox[16pt][l]{}} \\[-2pt] % review_12pt dist was added
$T,\left|\nabla T\right|,a_{\mathrm{t}}$ & $1.52$ \\[-2pt] % review_12pt dist was added
$T,\kappa,\left|\nabla T\right|,a_{\mathrm{t}}$ & $1.35$ \\[-2pt] % review_12pt dist was added
$T,\kappa,\left|\nabla T\right|,\frac{1}{\left|\nabla T\right|} \, \frac{\mathrm{D}_{T}\left( \left|\nabla T\right| \right) }{\mathrm{D}_{T} \left( t \right) }$ & $1.36$ \\
%\hline
%$T,\left|\nabla T\right|,\frac{1}{A}\frac{\mathrm{d}A}{\mathrm{d}t}$ & $1.60$ \\
%$T,\left|\nabla T\right|,\frac{1}{M}\frac{\mathrm{d}M}{\mathrm{d}t}$ & $1.64$ \\
\hline
\end{tabular}
\label{tab:irr_err}
\end{table} %%%%%%%%%%%%%%%%%%%%%%%%%%%%%
The optimal estimator analysis has quantitatively shown that at a given progress variable (temperature), preferential diffusion effects on the heat release rate are well represented by the parameter set $\left(\phi, h,Y_{\mathrm{H}}\right)$. {\color{black}To illustrate this finding,  the data that was shown in Fig.~\ref{fig:planar_jPDF_hr_vs_temp}\,(a,b) has been additionally conditioned on ${\left(\left<\left.\phi \,\right|T \right>, \left<\left.h \,\right|T \right>, \left<\left.Y_{\mathrm{H}} \,\right|T \right> \right)}$ for each flame, i.e. at a given temperature, all points share the same mixture state characterized by $\left(\phi, h,Y_{\mathrm{H}}\right)$. 
%, i.e. the remaining data points correspond to a fixed profile of all three mixture state parameters throughout the flame structure. 
As shown in Fig~\ref{fig:planar_jPDF_hr_vs_temp_cond}, the remaining scatter around the conditional mean heat release rate is similar for both ${\mathrm{Le}>1}$ and ${\mathrm{Le}=1}$, i.e. the differential-diffusion-induced heat release variations are contained in this reduced representation of the local mixture state.} The coupling between $\left(\phi, h,Y_{\mathrm{H}}\right)$ and the third-level (kinematic) parameters ($\kappa, \left|\nabla T\right|$) as well as the role of external effects (e.g.\ turbulence, cf.\ Fig.~\ref{fig:PrefDiff_schematic}), will be analyzed in Part~II of the present study~\cite{Falkenstein19_kernel_Le_II_cnf}. {\color{black}As a final note on the selection of suitable marker species, it should be mentioned that for a given~$Y_{\mathrm{H}_2}$ as a representative parameter for~$\phi$, there is some correlation between~$\left|\nabla T\right|$ (or $\chi$) and~$Y_{\mathrm{H}}$ (cf.\ Fig~\ref{fig:planar_jpdf_opt_est_spec}\,(b)), which according to Fig.~\ref{fig:PrefDiff_schematic} may explain the low irreducible error observed for the parameter set ${\left(T,Y_{\mathrm{H}_2},Y_{\mathrm{H}}\right)}$.} \par
\begin{figure} %%%%%%%%%%%%%%%%%%%%%%%%%%%%%
\centering
\begin{minipage}[b]{0.45\textwidth}
  \graphicspath{{./data/FLAME_PLANAR_DNS_02/190504_HR_condT_optEst/t_3.6E-04//}}
  \centering
%trim={<left> <lower> <right> <upper>}
%\hfill\vspace{0.1cm}
  %\makebox[0pt][l]{\quad(a)}\includegraphics[trim={10cm -4cm 40cm -6cm},clip,width=5.5cm]{}
\input{template/change_font_10.tex} % review_12pt
  \makebox[0pt][l]{\quad(a)}\scalebox{0.7}{\input{./data/FLAME_PLANAR_DNS_02/190504_HR_condT_optEst/t_3.6E-04/planar_dns_t3p6e-4_jPDF_HR_vs_T_condENTH_PHI_H_P02_ltx_JFM_HALF.tex}}
\input{template/change_font_12.tex} % review_12pt
\end{minipage}
\begin{minipage}[b]{0.45\textwidth}
  \graphicspath{{./data/FLAME_PLANAR_DNS_01/190504_HR_condT_optEst/t_3.6E-04//}}
  \centering
  % trim={<left> <lower> <right> <upper>}
  %\hfill\vspace{0.1cm}
\vspace{0.1cm}
\input{template/change_font_10.tex} % review_12pt
  \makebox[0pt][l]{\quad(b)}\scalebox{0.7}{\input{./data/FLAME_PLANAR_DNS_01/190504_HR_condT_optEst/t_3.6E-04/planar_dns_t3p6e-4_jPDF_HR_vs_T_condENTH_PHI_H_P01_ltx_JFM_HALF.tex}}
% Plot_jPDF_HR_vs_T_condENTH_PHI_H_P01_ltx_JFM.gp  jPDF_hr_temp_condENTH_PHI_H_gp.dat
\input{template/change_font_12.tex} % review_12pt
\end{minipage}
\vspace{0.1cm}
\caption{Planar Flame: Joint-PDFs of heat release rate and temperature for the ${\mathrm{Le}=1}$ dataset~(a) and the engine-relevant ${\mathrm{Le}>1}$ flame~(b) at ${t=2.8\,\tau_{\mathrm{t}}}$. {\color{black}Different from Fig.~\ref{fig:planar_jPDF_hr_vs_temp},} all data has been conditioned on ${\left(\left<\left.\phi \,\right|T \right>, \left<\left.h \,\right|T \right>, \left<\left.Y_{\mathrm{H}} \,\right|T \right> \right)}$. {\color{black}The reduced variance in subfigure~(b) compared to Fig.~\ref{fig:planar_jPDF_hr_vs_temp}\,(b) illustrates that heat release variations due to differential diffusion effects are contained in these three parameters.}}
\label{fig:planar_jPDF_hr_vs_temp_cond}
\end{figure} %%%%%%%%%%%%%%%%%%%%%%%%%%%%%
{\color{black}The preceding empirical parameter tests based on the parametric representation of the governing flame physics shown in Fig.~\ref{fig:PrefDiff_schematic} lead in fact to similar parameters on the third and fourth level as previously derived from the premixed flamelet equations by Savard and Blanquart~\cite{Savard17_nonLe_flamelet_model}. In the unity Lewis number limit, the scalar dissipation rate $\chi_c = 2 D_{\mathrm{th}} \left|\nabla c \right|^2$ was shown to be the only parameter in the flamelet equations, while for non-unity-Lewis-number flames, the diffusion rate of the progress variable $\xi_c = \frac{\partial }{\partial x_j}\left(\rho D_{\mathrm{th}} \frac{\partial c}{\partial x_j}\right)$ occurs as additional parameter. To relate the present findings to the optimal estimator tests performed by Savard and Blanquart, the following decomposition of the diffusion rate of the progress variable~$\zeta$ (cf.\ Eq.~(\ref{eq:c0_eq})) is considered:
\begin{align}
\xi_{\zeta} =	\frac{\partial}{\partial x_j} \left( \rho D_{\mathrm{th}} \frac{\partial \zeta }{\partial x_j} \right) & = -\rho D_{\mathrm{th}} \left|\nabla \zeta \right| \kappa_{\zeta} - \frac{\partial }{\partial x_{\mathrm{n}}}\left(\rho D_{\mathrm{th}} \left|\nabla \zeta \right| \right) \nonumber \\
& = \xi_{\zeta,\kappa} + \xi_{\zeta,{\mathrm{n}}}.
\label{eq:xi_zeta_eq}
\end{align}
Note that in contrast to the results presented in Tab.~\ref{tab:irr_err}, $\zeta$ is used here since diffusive transport is entirely contained in $\xi_{\zeta}$, which is not the case for temperature. From Eq.~(\ref{eq:xi_zeta_eq}), it can already be concluded that the tested parameter set $\left(T,\left|\nabla \zeta \right|,\kappa_{\zeta}\right)$ (or $\left(T,\left|\nabla T \right|,\kappa\right)$ in Tab.~\ref{tab:irr_err}) essentially represents $\xi_{\zeta,\kappa}$, while~$\xi_{\zeta,{\mathrm{n}}}$ has not yet been tested for the present datasets. In Tab.~\ref{tab:irr_err_savard}, optimal estimator results are shown for the third-level parameters known from Tab.~\ref{tab:irr_err}, but expressed in terms of gradients of $\zeta$ and complemented by~$\xi_{\zeta,\kappa}$ and~$\xi_{\zeta,{\mathrm{n}}}$. Overall, the trends observed in Tab.~\ref{tab:irr_err} hold for the progress variable as well, and the low irreducible error shown for $\left(T,\left|\nabla \zeta \right|,\xi_{\zeta,\kappa},\xi_{\zeta,\mathrm{n}}\right)$ confirms the effectiveness of the parameter combination~$\chi_c$  and~$\xi_{c}$, which was derived and tested for non-unity-Lewis-number flames by Savard and Blanquart~\cite{Savard17_nonLe_flamelet_model}. It should be noted that a strong correlation between the local heat release rate and~$\xi_{\zeta}$ can be expected from Eq.~(\ref{eq:c0_eq}), since the chemical source term is entirely balanced by the diffusion rate. Instead, the flame structure and geometry have been parameterized by~$\left|\nabla \zeta \right|$ and~$\kappa_{\zeta}$ in the present study, since the transport equations for both quantities highlight dependencies on external effects such as hydrodynamic strain. Replacing~$\left(\left|\nabla \zeta \right|,\kappa_{\zeta}\right)$ by~$\left(\left|\nabla \zeta \right|,\xi_{\zeta,\kappa}\right)$ gives almost identical irreducible errors, as shown in Tab.~\ref{tab:irr_err_savard}. However, the effectiveness of~$\xi_{\zeta,\mathrm{n}}$ in reducing the irreducible error suggests to analyze the role of external effects on this term in more detail in the future.} \par
\begin{table}%[!htb] %%%%%%%%%%%%%%%%%%%%%%%%%%%%%
\caption{{\color{black}Irreducible error intrinsic to heat release rate predictions from different input parameter sets (normalized by the temperature-based error value computed from the ${\mathrm{Le}=1}$ flame), similar to the study by Savard and Blanquart~\cite{Savard17_nonLe_flamelet_model}. $\xi$ is defined in Eq.~(\ref{eq:xi_zeta_eq}). Data in parentheses corresponds to the ${\mathrm{Le}=1}$ flame.}}
\vspace{0.1cm}
\centering
%\begin{tabular}{P{.22\textwidth}|P{.22\textwidth}}
\begin{tabular}{P{.35\textwidth}|P{.35\textwidth}} %% review_12pt
\hline
%\\ [-10pt] % review_12pt
Parameter Set $\Pi$ & $\epsilon^2_{\mathrm{Le}>1}\left(\Pi\right) \;/\; \epsilon^2_{\mathrm{Le}=1}\left(T\right) $  \\[-2pt] % review_12pt dist was added
%\\ [-10pt] % review_12pt
\hline
$T,\kappa_{\zeta}$ & $4.23$ \\[-2pt] % review_12pt dist was added
%$T,\left|\nabla \zeta \right|$ & $1.64$ \\[-2pt] % review_12pt dist was added
$T,\left|\nabla \zeta \right|$ & \multicolumn{1}{r}{$1.64$\quad($0.63$)\makebox[16pt][l]{}} \\[-2pt] % review_12pt dist was added
$T,\left|\nabla \zeta \right|,\kappa_{\zeta}$ & $1.38$ \\[-2pt] % review_12pt dist was added
$T,\left|\nabla \zeta \right|,\xi_{\zeta,\kappa}$ & $1.37$ \\[-2pt] % review_12pt dist was added
$T,\left|\nabla \zeta \right|,\xi_{\zeta,\mathrm{n}}$ & $1.17$ \\[-2pt] % review_12pt dist was added
$T,\left|\nabla \zeta \right|,\kappa_{\zeta},\xi_{\zeta,\mathrm{n}}$ & $0.93$ \\[-2pt] % review_12pt dist was added
$T,\left|\nabla \zeta \right|,\xi_{\zeta,\kappa},\xi_{\zeta,\mathrm{n}}$ & $0.92$ \\[-2pt] % review_12pt dist was added
%$\left(\mathrm{Le}=1:\; T,\left|\nabla \zeta\right|\right)$ & $\left( 0.63\right)$ \\[-2pt] % review_12pt dist was added
\hline
\end{tabular}
\label{tab:irr_err_savard}
\end{table} %%%%%%%%%%%%%%%%%%%%%%%%%%%%%
{\color{black}In order to use the results provided in Tab.~\ref{tab:irr_err} or~\ref{tab:irr_err_savard} for the design of a reduced-order chemistry model, e.g. based on flamelet generated manifolds (FGM)~\cite{Oijen00_fgm}, several additional considerations will be necessary. The preceding analysis may be considered as a first step that resulted in quantitative criteria for selecting a viable parameterization for the heat-release-rate response to differential diffusion from the tested parameter set candidates. Additionally, the control variable space to be selected should provide a unique mapping to the FGM, which would need to be assessed in a second step. Knudsen et al.~\cite{Knudsen13_strained_model} tested several control variable candidates in this regard and identified non-unique mapping as an issue for the parameters $Y_{\mathrm{H_2}}$, $Y_{\mathrm{CO}}$ (first-level), as well as the scalar dissipation rate $\chi_c$ (third-level) in a strained methane/air flame, particularly at high strain rates. Third, the (conditional) controlling variables or relevant statistical moments need to be determined from the reacting flow simulation (DNS, LES, RANS), which may introduce a need for additional closure models and numerical methods. In this respect, it will be more challenging to develop a model that, e.g. relies on a transport equation for $\chi_c$ (third-level)~\cite{Swaminathan05_chi_eq} than using a transport equation for mixture fraction (representative for local equivalence ratio, second-level)~\cite{Regele13_zmix_premixed_model,Schlup19_zmix_premixed_model}, or tracking a species mass fraction such as~$Y_{\mathrm{H}}$ (first-level)~\cite{Knudsen13_strained_model}. For LES, $\chi_c$ may instead be computed from an eddy-diffusivity-based time scale and an appropriately modeled subfilter scalar variance, in analogy to previous work on non-premixed combustion modeling~\cite{Raman05_chi_algebr}. In previous modeling studies on stretched premixed flames, the FGM was typically extended by one additional controlling variable in order to account for differential diffusion effects. Regarding first-level parameters, $Y_{\mathrm{CO}}$~\cite{vanOijen02_premixedCntFlow_Le}, $Y_{\mathrm{OH}}$~\cite{vanOijen07_strain_curv_fgm} and $Y_{\mathrm{H}}$~\cite{Knudsen13_strained_model} were considered, but mainly in methane/air flames with effective Lewis numbers close to unity. As shown in this study, for modeling of flames with Lewis numbers significantly different from unity, (additionally) selecting a species that correlates with the local equivalence ratio might be more suitable. Second-level parameters were mostly used for laminar~\cite{Regele13_zmix_premixed_model} or resolved-turbulence simulations~\cite{vanOijen16_fgm_precs,Schlup19_zmix_premixed_model} with reduced chemistry parameterized by mixture fraction or a combination of element mass fractions. Here, it should be noted that changes in element mass fraction and enthalpy may decouple, i.e. require more than one additional FGM controlling variable at high-stretch conditions~\cite{vanOijen02_premixedCntFlow_Le}. Concerning third-level parameters, $\chi_c$ was used in RANS~\cite{Kolla10_strained_flamelet_model} and LES~\cite{Thornber11_chi_les} studies. With respect to fourth-level parameters it should be noted that the addition of a dynamic parameter in the optimal estimator analysis was motivated by capturing transient effects, which are not covered by steady FGM approaches per se. Additionally, the dynamic parameter would need to be unambiguously defined in turbulent and canonical laminar flame configurations~\cite{Knudsen13_strained_model}. }\par
\begin{figure}
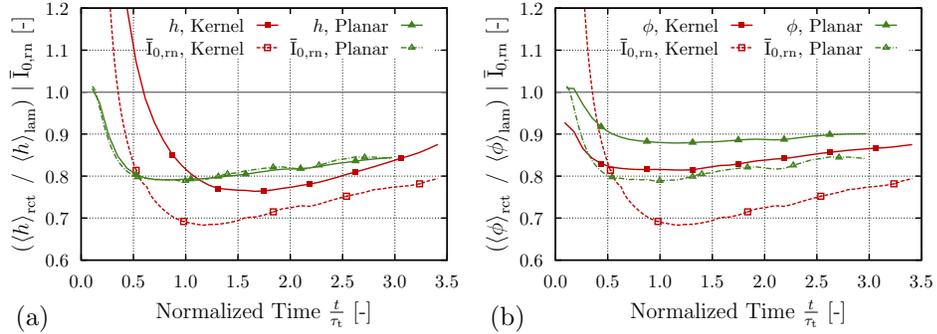
 %%%%%%%%%%%%%%%%%%%%%%%%%%%%%
\centering
\begin{minipage}[b]{0.45\textwidth}
  \graphicspath{{./data/FLAME_KERNEL_DNS_01/190330_local_equiv_ratio_avg_rct_zone/}}
  \centering
%trim={<left> <lower> <right> <upper>}
%\hfill\vspace{0.1cm}
%  \makebox[0pt][l]{\quad(a)}\includegraphics[trim={0cm -0.75cm 0cm 0cm},clip,width=7cm]{./data/FLAME_PLANAR_DNS_01/190316_local_equiv_ratio_curv/t_3.600E-04/}
\input{template/change_font_10.tex} % review_12pt
  \makebox[0pt][l]{\quad(a)}\scalebox{0.7}{\input{./data/FLAME_KERNEL_DNS_01/190330_local_equiv_ratio_avg_rct_zone/kernel_dns_avg_Enthalpy_rct_zone_I0_time_01_P01_JFM_HALF.tex}}
% Plot_Enthalpy_rct_zone_AVG_time_01_P01_ltx_JFM.gp
\input{template/change_font_12.tex} % review_12pt
\end{minipage}
\begin{minipage}[b]{0.45\textwidth}
  \graphicspath{{./data/FLAME_KERNEL_DNS_01/190330_local_equiv_ratio_avg_rct_zone/}}
  \centering
  % trim={<left> <lower> <right> <upper>}
  % \hfill\vspace{0.1cm}
\input{template/change_font_10.tex} % review_12pt
    \makebox[0pt][l]{\quad(b)}\scalebox{0.7}{\input{./data/FLAME_KERNEL_DNS_01/190330_local_equiv_ratio_avg_rct_zone/kernel_dns_avg_local_equiv_ratio_rct_zone_I0_time_01_P01_JFM_HALF.tex}}
% Plot_local_equiv_ratio_rct_zone_AVG_time_01_P01_ltx_JFM.gp
\input{template/change_font_12.tex} % review_12pt
\end{minipage}
\caption{Reaction-zone-averaged enthalpy (a) and local equivalence ratio (b) {\color{black}compared to the reaction/normal diffusion stretch factor~$\overline{\mathrm{I}}_{0,\mathrm{rn}}$} as function of time for ${\mathrm{Le}>1}$.  Note that both~$h$ and~$\phi$ are approximately constant in ${\mathrm{Le} = 1}$ flames.}
\label{fig:kernel_planar_h_phi_vs_time}
\end{figure}

% ToDo: say something that CH4 flames are not representative. a marker species may need to include the effect of mixture state and time scale (Ed: YH-> probably time scale due to Le ~ 1). Z/h may not respond to time scale. chi may include both. YH2 maybe also.
To finally connect the presented macroscopic and micro-scale analyses, the correlation between the local mixture state parameters~$h$ and~$\phi$ with~$\overline{\mathrm{I}}_{0,\mathrm{rn}}$  (cf.\ Fig.~\ref{fig:kernel_I0_Sigma}\,(a)) during flame development will be briefly discussed hereafter. {\color{black}For the purpose of this analysis, both $h$ and~$\phi$ have been averaged over the reaction zones of the ${\mathrm{Le}>1}$ flames. The time evolution of both parameters and $\overline{\mathrm{I}}_{0,\mathrm{rn}}$ is shown in Fig.~\ref{fig:kernel_planar_h_phi_vs_time}. For the planar flame configuration, $\overline{\mathrm{I}}_{0,\mathrm{rn}}$ almost exactly follows the history of the normalized reaction zone enthalpy. Here, the reference enthalpy for normalization was obtained by applying the reaction zone averaging in an unstretched laminar flame. 
%, which is analogous to the burning velocity normalization used for the stretch factor definition. 
In the flame kernel configuration, the initial decrease of reaction zone enthalpy from the initial conditions due to the development of differential diffusion effects observed in the planar flame is effectively delayed by spark ignition. While the flame kernel stretch factor qualitatively follows the normalized reaction zone enthalpy evolution, both lines deviate significantly. This is likely due to the lean mixture conditions that rapidly develop inside the reaction zone of the flame kernel (cf.\ Fig.~\ref{fig:kernel_planar_h_phi_vs_time}\,(b)). At ${t\approx0.55\,\tau_{\mathrm{t}}}$,  $\overline{\mathrm{I}}_{0,\mathrm{rn}}$ is approximately equal in both flame configurations. At this time instant, the flame kernel exhibits noticeable excess enthalpy, but substantially leaner mixture inside the reaction zone as compared to the planar flame. When the ignition enthalpy decays, the burning velocity of the flame kernel reaches a minimum and then slowly recovers, in line with the mixture state as the positive global mean curvature reduces during flame kernel growth.} \par 
%Note that the planar flame was initialized as a laminar unstretched flame, which features slightly rich mixture in the reaction zone. By contrast, the flame kernel evolves from a pocket of excess enthalpy. Consequently, the flame kernel enthalpy drops below the enthalpy level of the planar flame not before ${t=1.2\,\tau_{\mathrm{t}}}$ as shown in Fig.~\ref{fig:kernel_planar_h_phi_vs_time}\,(a), while the mixture composition in both flames approaches the minimum equivalence ratio level already at ${t=0.6\,\tau_{\mathrm{t}}}$ (cf.\ Fig.~\ref{fig:kernel_planar_h_phi_vs_time}\,(b)). The combination of both effects influences the local burning velocity evolution (cf.\ Fig.~\ref{fig:kernel_I0_Sigma}\,(a)). 
% (cf.\ Fig.~\ref{fig:kernel_I0_Sigma}\,(b))
%Regarding differences between flame geometries, the flame kernel shows a minimum local equivalence ratio well below the stationary level of the planar flame, which is expected due to the positive mean curvature. \par
%
 %%%%%%%%%%%%%%%%%%%%%%%%%%%%%
Since small flame kernels are particularly sensitive to external perturbations, the reduction in burning velocity (cf.\ Fig.~\ref{fig:kernel_I0_Sigma}\,(a)) 
caused by low enthalpy and equivalence ratio inside the reaction zone in ${\mathrm{Le}>1}$ mixtures is critical in terms of~CCV in engines. However, systematic development of an appropriate spark ignition strategy may enable effective counter measures. From laminar investigations it is known that the minimum ignition energy (MIE) increases with increasing Lewis number~\cite{Tromans88_1d_dns_mie,Chen11_spark_ign_energy}, which is in line with the present turbulent flame results. \par
%
%============================================================================================================================================
\section*{Conclusions}
\label{sec:conclusions}
The intention of the present work is to clarify the role of differential diffusion during early flame kernel development in the thin reaction zones regime. Due to the expected relevance for the occurrence of cycle-to-cycle variations in SI engines, a DNS database has been carefully designed to be representative of practical engine conditions. Conclusive analyses were enabled by systematic variations of the global flame geometry and the mixture Lewis number of a realistic transportation fuel and air. A macroscopic analysis of the global burning rate as practical quantity of interest has shown the impact of differential diffusion during early flame kernel development. Further, the local heat release rate response has been used as starting point for a systematic micro-scale analysis on the complex parameter dependencies inside the turbulent flame structure. 
%The results have been related to canonical turbulent and laminar flame configurations to enhance generality and increase the relevance of the present findings for future modeling efforts. 
Specifically, the following conclusions can be drawn: 
\begin{itemize}
\item The detrimental effect of differential diffusion on the net burning rate of flame kernels {\color{black} with fuel ${\mathrm{Le}>1}$} is mainly due to changes in the normal propagation velocity, while differences in flame area at the same flame radius are slightly less important, though still significant under the present conditions. This is in agreement with the findings of Dinesh et al.~\cite{Dinesh16_spherical_H2_dns} in flames with ${\mathrm{Le}<1}$. However, the observation that the flame-area dependence on thermal-diffusive effects is weakened under high-Reynolds-number flow conditions~\cite{Dinesh16_spherical_H2_dns} may not translate to the very early phase of flame kernel development, which is a topic suggested for future work.
\item Early flame kernel area growth is significantly reduced in the ${\mathrm{Le}>1}$ flame, which has been attributed to a reduction in the normal-propagation term in the surface area evolution equation. Area production by propagation of the curved flame surface is proportional to the normal-propagation stretch factor $\overline{\mathrm{I}}_{0,\mathrm{rn}}$, which has been shown to strongly suffer from low enthalpy and local equivalence ratio levels inside the reaction zone as a consequence of differential diffusion. Under the present turbulent conditions, differential diffusion thus acts on initial flame area growth in a very similar way as known from laminar spherical flames.
\item Non-unity Lewis numbers were shown to result in strong variations in local heat release rate. In particular, very low heat release rates were locally observed. This is in agreement with results reported by Shim et al.~\cite{Shim13_h2_dns_hrr_fluct_thin_rct_zones} for $\mathrm{H}_2$/air flames located in the thin reaction zones regime. However, such locally low heat release rates were neither observed under flame conditions in the corrugated flamelets regime by Shim et al., nor in the present ${\mathrm{Le}=1}$ datasets in the thin reaction zones regime. Hence, heat release fluctuations seem to be particularly pronounced in ${\mathrm{Le}\neq1}$ flames located in the thin reaction zones regime. Both of these conditions are typically found in SI engines. {\color{black}The present results on iso-octane/air flames are in agreement with previous work on higher hydrocarbon fuels~\cite{Savard15_Le_effects_C7H16_highKa,Lapointe15_appendix_avg}.}
% Savard15:~\cite{Savard15_Le_effects_C7H16_highKa} However, large source term fluctuations were present in both flames and in particular, local extinction events were found for the flame with non-unity Lewis numbers. This showed that a reaction zone can at the same time be thin and broken.
% Lapointe15: {Lapointe15_appendix_avg}  Third, local extinctions and chemical source terms fluctuations were investigated and quantified through probability density functions at the temperature of peak reactions. Large fluctuations in fuel consumption rate were observed for both non-unity and unity Lewis number simulations, increasing with the reaction zone Karlovitz number. However, it is unclear, especially in the context of LES filtering, if these chemical source term fluctuations need to be incorporated in low-order models. These large fluctuations only lead to local extinction events in the non-unity Lewis number flames, and the probability of these events decreased at high Karlovitz numbers. In fact, less than 1% of the flame surface underwent extinction at Kad > 10. This was explained by the competing effects of increased curvatures and decreased differential diffusion effects as the reaction zone Karlovitz number was increased.
%
\item By using the optimal estimator concept, the fluctuations in heat release rate for a given temperature (or progress variable) have been quantitatively attributed to different parameter dependencies and related to the behavior of a ${\mathrm{Le}=1}$ flame. Specifically, a hierarchical grouping of parameter sets based on coupling-strength with the quantity of interest (heat release rate) has been proposed, i.e.\ first-level ({\color{black}temperature and species mass fractions,} reaction/diffusion-controlled), second-level ({\color{black}equivalence ratio and enthalpy,} diffusion-controlled), third-level ({\color{black}curvature and scalar gradient magnitude,} kinematic) and fourth-level ({\color{black}tangential strain as a measure for time-rate-of-change of third-level parameters,} kinematic/dynamic) parameters. On each level, a dominant parameter can be identified (cf.\ Tab.~\ref{tab:irr_err}), which may be useful for modeling purposes. For the present analysis, the objective was to find a reduced representation of the local mixture state, which determines the heat release rate. 
%On the first level, $Y_{\mathrm{H}_2}$ is the dominant parameter, while on the second level, the local equivalence ratio~$\phi$ turned out to be most representative. The third level is best described by $\left|\nabla T \right|$. Accounting for variations in both local equivalence ratio and enthalpy reduced the conditional variance in heat release rate below the value observed in the unconditional ${\mathrm{Le}=1}$ flame, while $\epsilon^2\left(T, \phi, h,Y_{\mathrm{H}}\right)$ was found to be the lowest irreducible error among all tested parameter sets. 
It turned out that conditioning on the parameter set ($T, \phi, h,Y_{\mathrm{H}}$) reduces the remaining scatter around the conditional mean heat release rate to identical levels in both ${\mathrm{Le}>1}$ and ${\mathrm{Le}=1}$  flames, which is in agreement with the proposed flame physics schematic based on findings from literature (cf.\ Fig.~\ref{fig:PrefDiff_schematic}). %\textcolor{red}{ToDo: revise according to CNF\_kernel\_Le\_effect\_190712\_HP\_02\_TF\_clean.pdf}
%
%\item Besides positive curvature, the flame structure is detrimentally altered by hydrodynamic strain. In the limit of unity Lewis number, the production of scalar gradients by heat release becomes mostly independent on the scalar gradient field itself, which makes the turbulent flame almost insensitive to gradient changes by hydrodynamic strain (cf.\ std. dev. in Fig.~\ref{fig:planar_hr_vs_temp}\,(b)).
\end{itemize}
The coupled nature of the local mixture state with the flame geometry and structure due to differential diffusion is discussed in Part~II of the present study~\cite{Falkenstein19_kernel_Le_II_cnf}.
\section*{Acknowledgement}
The authors from RWTH Aachen University gratefully acknowledge partial funding by Honda R\&D and by the Deutsche Forschungsgemeinschaft (DFG, German Research Foundation) under Germany's Excellence Strategy - Exzellenzcluster 2186 `The Fuel Science Center' ID: 390919832.\par
%The authors from RWTH Aachen University gratefully acknowledge partial funding by Honda R\&D and by the Cluster of Excellence “Tailor-Made Fuels from Biomass,” which is funded by the Excellence Initiative of the German federal and state governments to promote science and research at German universities.\par
The authors gratefully acknowledge the Gauss Centre for Supercomputing e.V. (www.gauss-centre.eu) for funding this project by providing computing time on the GCS Supercomputer Super-MUC at Leibniz Supercomputing Centre (LRZ, www.lrz.de).\par
Data analyses were performed with computing resources granted by RWTH Aachen University under project thes0373. \par
S.K. gratefully acknowledges financial support from the National Research Foundation of Korea (NRF) grant by the Korea government (MSIP) (No. 2017R1A2B3008273). \par
T.F. would like to thank G{\"u}nter Paczko,  sadly no longer with us, for the helpful discussions in all these years.
\appendix
\setcounter{figure}{0}
%-------------------------------------------------------------------------------------------------------------------------------------------
\section{Correlation between Marker Species and Mixture State Parameters}
\label{apx:marker_spec_corr}
{\color{black}The optimal estimator results presented in Tab.~\ref{tab:irr_err} suggest to parameterize the local mixture state with the quantities $\left(\phi, h,Y_{\mathrm{H}}\right)$ for a given temperature. Further, some parameters that were assigned to different hierarchical groups in fact showed similar performance. Specifically, the first-level parameter set $\left(T,Y_{\mathrm{H}_2}\right)$ was shown to yield an almost identical irreducible error as the second-level parameter set $\left(T,\phi\right)$. Similarly, the first-level parameter set $\left(T,Y_{\mathrm{H}}\right)$ and the third-level parameter set $\left(T,\left|\nabla T \right|\right)$ yielded comparable irreducible errors. Here, the correlation of these marker species with the representative mixture state ($\phi$) and flame structure ($\left|\nabla T \right|$) parameters shall be assessed. The joint-PDFs shown in Fig.~\ref{fig:planar_jpdf_opt_est_spec} confirm a correlation between~$\phi$ and~$Y_{\mathrm{H}_2}$ for a given temperature and fixed~$Y_{\mathrm{H}}$, as well as a correlation between~$\left|\nabla T \right|$ and~$Y_{\mathrm{H}}$ for a given temperature and fixed~$Y_{\mathrm{H}_2}$. Hence, $Y_{\mathrm{H}_2}$ can be considered as a marker for the differential-diffusion effect on the local mixture composition, while ~$Y_{\mathrm{H}}$ may contain information on the local mixture internal energy distribution caused by changes in flame structure (cf.\ Fig.~\ref{fig:PrefDiff_schematic}).}

\begin{figure} %%%%%%%%%%%%%%%%%%%%%%%%%%%%%
\centering
\begin{minipage}[b]{0.45\textwidth}
  \graphicspath{{./data/FLAME_PLANAR_DNS_01/190316_local_equiv_ratio_curv/t_3.600E-04/}}
  \centering
%trim={<left> <lower> <right> <upper>}
%\hfill\vspace{0.1cm}
\input{template/change_font_10.tex} % review_12pt
  \makebox[0pt][l]{\quad(a)}\scalebox{0.7}{\input{./data/FLAME_PLANAR_DNS_01/190316_local_equiv_ratio_curv/t_3.600E-04/planar_dns_t3p6e-4_jPDF_PHI_vs_H2_P01_ltx_JFM_HALF_200216.tex}}
\input{template/change_font_12.tex} % review_12pt
% Plot_jPDF_HR_vs_Enth_cond_TmaxHR_PHI_H_P01_ltx_JFM.gp jPDF_hr_enth_condT_maxHR_condPHI_H_gp.dat main_only_plots_HR_Enth_jPDF_cond_condPHI_H.m
\end{minipage}
\begin{minipage}[b]{0.45\textwidth}
  \graphicspath{{./data/FLAME_PLANAR_DNS_01/190316_local_equiv_ratio_curv/t_3.600E-04//}}
  \centering
  % trim={<left> <lower> <right> <upper>}
\vspace{0.1cm}
\input{template/change_font_10.tex} % review_12pt
  \makebox[0pt][l]{\quad(b)}\scalebox{0.7}{\input{./data/FLAME_PLANAR_DNS_01/190316_local_equiv_ratio_curv/t_3.600E-04/planar_dns_t3p6e-4_jPDF_gradT_vs_H_P01_ltx_JFM_HALF_200216.tex}}
\input{template/change_font_12.tex} % review_12pt
\end{minipage}
\vspace{0.1cm}
\caption{\color{black}Planar Flame: Joint-PDF of the local equivalence ratio~$\phi$ conditioned on ${\left(T=T_{\mathrm{max}Y_{\mathrm{H}_2}},\left<\left.Y_{\mathrm{H}}\, \right|T=T_{\mathrm{max}Y_{\mathrm{H}_2}} \right>\right)}$ and the mass fraction of molecular hydrogen~$Y_{\mathrm{H}_2}$~(a); of the temperature gradient magnitude~$\left|\nabla T \right|$ conditioned on ${\left(T=T_{\mathrm{max}Y_{\mathrm{H}}},\left<\left.Y_{\mathrm{H}_2}\, \right|T=T_{\mathrm{max}Y_{\mathrm{H}}} \right>\right)}$ and the H-radical mass fraction~$Y_{\mathrm{H}}$. Here, the data have been conditioned on temperature iso-surfaces corresponding to the maximum species mass fraction in the laminar unstretched flame.}
\label{fig:planar_jpdf_opt_est_spec}
\end{figure} %%%%%%%%%%%%%%%%%%%%%%%%%%%%%
%
%\section{ToDo}
%
%- Find the final representation of Fig.~\ref{fig:PrefDiff_schematic}. \par
%- Can we quantify Markstein effects and separate them from the total stretch effect? Could condition planar-flame data on weak stretch (say Ka lt 0.1 or phi and h) and use my spark model implementation to predict Markstein length. Then check joint pdf of model (rho sd) vs. DNS (rho sd). Check JFM Thiesset? \par
%- check all slides again \par
%- find good nomenclature for unity-Le-number flames etc \par
%- make equations in Fig 1 consistent \par
%- check mass burned / transition to turbulent flame again in slide. how much is the transition delayed by differential diffusion? criterion should work on one kernel realization. \par
%
%\par
%\textbf{Merits of this paper} \par
%- While several studies have highlighted the mechanisms/terms dominating some aspects of the overall cause-effect chain, the interplay of the governing parameters in a semi-realistic setting for a practical combustion device has not been studied. (check introduction Bariki) \par
%- Comprehensive/systematic application-driven analysis of differential diffusion effects \par
%- hierarchical parameter dependencies of the heat release rate are presented using the optimal estimator concept \par
%- flame kernel is related to more canonical configuration of a planar flame, which is shown to on average behave as a laminar flame (laminar flame structure is well-described by Pitsch). Useful for modeling. \par
%- Quantitative assessment of flame structure under engine part-load conditions (with complex chemistry!) \par
%%%%%%%%%%%%%%%
\FloatBarrier
\bibliography{./references/flame_kernel_01,./references/03_b_Publikationen_anderer.bib,./references/les4ice16.bib,./references/rezchikova.bib}

\end{document}